\documentclass[11pt,a4paper,twocolumns]{revtex4}
\usepackage{amsmath}
\usepackage{graphicx}
\usepackage{fancyhdr}
\usepackage{calc}
\usepackage{amssymb}
\usepackage[sort&compress]{natbib}
\usepackage{setspace}
\usepackage{amsfonts}
\usepackage{commath}
\usepackage[innercaption]{sidecap}

\expandafter\def\expandafter\normalsize\expandafter{%
    \normalsize
    \setlength\abovedisplayskip{0pt}
    \setlength\belowdisplayskip{5pt}
    \setlength\abovedisplayshortskip{0pt}
    \setlength\belowdisplayshortskip{5pt}
}

\begin{document}


\setlength\arraycolsep{12pt}
\renewcommand{\arraystretch}{1.5}

\newcommand{\av}[1]
{
\langle #1 \rangle
}

\newcommand{\R} 
{
\mathbb R
}

\newcommand{\E} 
{
\mathbb E
}

\newcommand{\M} 
{
\mathbb M
}

\renewcommand{\P} 
{
\mathbb P
}

\newcommand{\B} 
{
\mathbb B
}

\newcommand{\m}[3]
{
#1_{#2 #3}
}

\newcommand{\D}[2]
{
\mathcal{D}_{#1 #2}
}

\renewcommand{\L}[2]
{
\mathcal{L}{(#2 \rightarrow #1)}
}

\newcommand{\T}[2]
{
T (#2 \rightarrow #1)
}

\title{Predicting the patterns of spatio-temporal signal propagation \\ in complex networks}

\author{Chittaranjan Hens$^{1}$, Uzi Harush$^{1}$, Reuven Cohen$^{1}$ \& Baruch Barzel$^{1}$}
\affiliation{
\begin{enumerate}
\item
Department of Mathematics, Bar-Ilan University, Ramat-Gan 52900, Israel 
\end{enumerate}
}

\maketitle

\vspace{-15mm}

{\bf
A major achievement in the study of complex networks is the observation that diverse systems, from sub-cellular biology \cite{Barabasi2004,Jeong2000,Alon2006} to social networks \cite{Palla2005,Boccaletti2006,Girvan2002}, exhibit universal topological characteristics \cite{DSouza2009,Achlioptas2009,Caldarelli2007,Drogovtsev2003,Strogatz2001,Helbing2008,Newman2010,PastorSatorras2004}. Yet this universality does not naturally translate to the dynamics of these systems \cite{Barrat2008,Holter2001,Strogatz2000,Arenas2008}, hindering our progress towards a general theoretical framework of network dynamics. The source of this theoretical gap is the fact that the behavior of a complex system cannot be uniquely predicted from its topology, but rather depends also on the dynamic mechanisms of interaction between the nodes \cite{Barzel2009}, hence systems with similar structure may exhibit profoundly different dynamic behavior. To bridge this gap, we derive here the patterns of network information transmission, indeed, the essence of a network's behavior \cite{Toroczkai2004,Borgatti2005,Vespignani2012}, by offering a systematic translation of topology into the actual spatio-temporal propagation of perturbative signals. We predict, for an extremely broad range of nonlinear dynamic models, that the propagation rules condense around three highly distinctive dynamic universality classes, characterized by the interplay between network paths, degree distribution and the interaction dynamics. Our formalism helps us leverage the major advances in the mapping of real world networks, into predictions on the actual dynamic propagation, from the spread of viruses in social networks \cite{PastorSatorras2015,Brockmann2013,Barthelemy2004,Barthelemy2005,Lloyd2001} to the diffusion of genetic information in cellular systems \cite{Endy2001,Maslov2007}.   
}

The spread of information in a complex system is mediated by its underlying topology, with the metric of network paths commonly assumed to be the main determinant of the propagation \citep{Maslov2007,Brockmann2013,Noh2004,Maayan2005}. This rationale has motivated a widespread effort to retrieve the structure of many real world networks \cite{Watts1998,Barabasi2002,Cohen2010}, which in turn emerged as a powerful tool to visualize and predict information propagation, such as epidemic spreading via air-traffic \cite{Brockmann2013,Balcan2009} or neuronal activity patterns along the pathways of the connectome \cite{Kumar2010}. In all these cases, the network topology exposes the {\it natural} geometry of the propagation, with network distance being the main predictor of the spreading behavior. Yet, network topology does not always capture information propagation in such a transparent fashion, due to the diverse forms of nonlinear interactions that may take place between the nodes \cite{Barzel2013,Barzel2015}. Indeed, as we demonstrate below, slight modifications in the system's dynamics can potentially have a profound impact on the observed propagation, causing similar networks to exhibit fundamentally different spreading patterns. This results in a seemingly unpredictable {\it zoo} of highly irregular propagation patterns, severely limiting our ability to systematically translate network topology into spatio-temporal propagation. Therefore, our goal here is to (i) \textit{expose} the potential propagation patterns; (ii) \textit{derive}, analytically, the rules that predict these observed patterns and (iii) \textit{translate}, based on our theoretical findings, the network topology into its predicted spatio-temporal spread, thus strengthening networks as the predictive tool of information propagation they are designed to be.

\subsection*{Observing signal propagation}

\begin{figure}[h!]
\centering
\includegraphics[angle=0,width = 14cm]{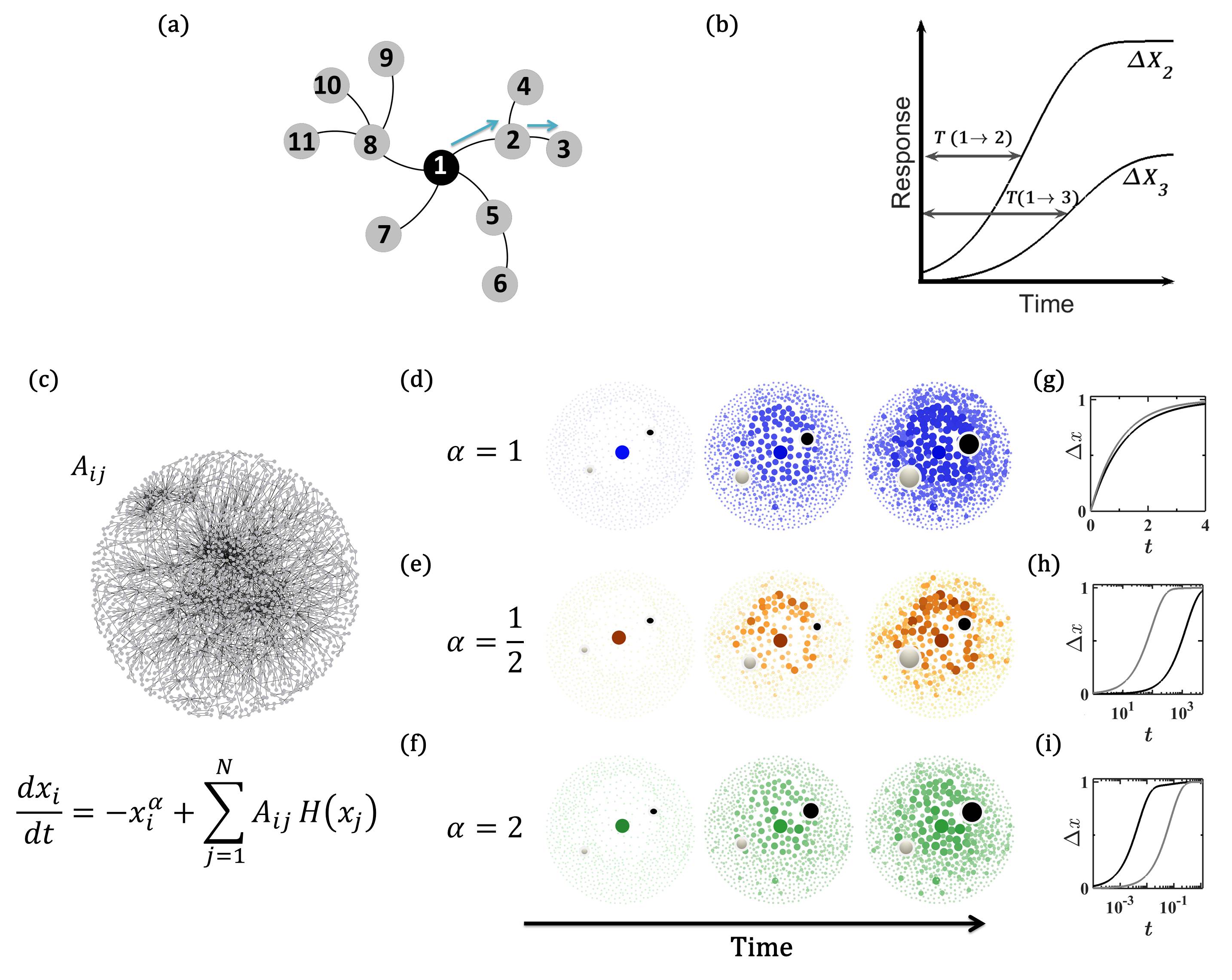}
\caption{{\bf Propagation of signals in a complex networks}.
The dynamic behavior of a complex network is captured by its patterns of information, or signal, propagation.
(a)
A local signal in the form of an activity perturbation $\Delta x_1$, applied on the source node $1$ (black) spreads through the network, impacting all other nodes $2,3,\dots$.
(b)
This spatio-temporal propagation is captured by the response $\Delta x_i(t)$, here depicted for nodes $2$ and $3$. The propagation time $\T ij$ captures the time in which $\Delta x_i$ reaches an $\eta$-fraction of its final response, here illustrated for $\eta = 1/2$ (half-life).
(c)
To model network dynamics we use a two layer description, exemplified here on the human protein-protein interaction network \cite{Rual2005}. The first layer is the topology $\m Aij$ (top). The second layer is the system's dynamics (equation, bottom), designed to capture the inner mechanisms driving the system's observed behavior. Here proteins are depleted at a rate $x_i^{\alpha}$ and activated by their neighboring proteins via the Hill function $\mathcal{H}(x_j) = x_j^h/(1 + x_j^h)$, where we set $h = 1$ \cite{Alon2006,Karlebach2008}.
(d)
Signal propagation as obtained for $\alpha = 1$. The response $\Delta x_i(t)$ is represented by the node's size and color depth. 
(e) - (f)
Changing the value of $\alpha$ impacts the propagation patterns, showing that $\m Aij$ alone is insufficient to predict information spread.
(g)
The temporal response, $\Delta x$ vs.\ $t$, of two specific nodes, marked in black and grey in the network layouts. For $\alpha = 1$ these two nodes exhibit a synchronous response, namely the signal reaches both at approximately the same time.
(h)
When $\alpha = 1/2$, however, the same two nodes receive the signal at different times, with grey responding approximately $10^2$ times earlier than black.
(i)
The sequence of responses is reversed for $\alpha = 2$, as now black responds before grey. Hence, minor changes in the dynamic equation in (c), may have profound and unpredictable consequences on the propagation.  
}
\label{Fig1}
\end{figure}

\noindent
To illustrate the challenge we begin with a specific example, using the human protein-protein interaction network \cite{Rual2005} to track the the propagation of biochemical signals in a sub-cellular environment. Denoting the abundance of the $i$th protein by $x_i(t)$, we can capture the system's dynamics through \cite{Voit2000} $\dot x_i = -B_i x_i^{\alpha} + \sum_{j = 1}^N \m Aij \mathcal H(x_j)$, in which the first term describes a protein's self-dynamics, \textit{e.g.}, degradation ($\alpha = 1$), dimerization ($\alpha = 2$) or a more complicated chain reaction (fractional $\alpha$, \cite{Laidler1987}), and the second term depicts $i$'s regulation by its interacting partners, often captured by a Hill function of the form \cite{Karlebach2008} $\mathcal{H}(x_j) = x_j/(1 + x_j)$; $\m Aij$ is the protein interaction network (Fig.\ \ref{Fig1}c). Changes in the abundance of one protein propagate, through $\m Aij$, to affect the abundance levels of all other proteins, representing a spread of biochemical information in the system \cite{Maslov2007}. Hence we initiate a biochemical \textit{signal} by introducing a perturbation $\Delta x_j$ to the steady state abundance of the {\it source} $j$, and then track its propagation, as it penetrates the network, to impact all {\it target} nodes $i = 1,\dots,N$ (Fig.\ \ref{Fig1}a,b). This process is analogous to, \textit{e.g.}, an \textit{over-expression} perturbation, a common procedure to track the spread of sub-cellular information \cite{Kauffman2004}.  

In Fig. \ref{Fig1}d-f we visualize this propagation, in selected time points, as obtained for three different values of the parameter $\alpha$. The signal source is at the center of each layout, and the response of all target nodes is represented by their size and color depth, hence proteins that receive the signal at earlier times appear first in each snapshot. We find that the patterns of propagation are highly irregular, with the signal appearing in different locations, depending on the system's dynamics ($\alpha$). For instance, in case $\alpha = 1$ the signal propagates, roughly, from the center to the periphery (blue), a rather intuitive form of propagation. Yet, on the same network, a slight modification of the dynamics ($\alpha = 1/2$) leads to different behavior, as now the signal seems to \textit{skip} the most adjacent nodes and appear first at more distant neighbors (red). To deepen our observation of the different response patterns, we focus on a specific pair of target nodes, highlighted in grey and black. In case $\alpha = 1$ (blue) we find that these two nodes exhibit similar behavior, featuring an almost synchronous response (Fig.\ \ref{Fig1}g). The picture dramatically changes, however, when $\alpha = 1/2$ (red), in which case the signal impacts the black node at a much later time (Fig.\ \ref{Fig1}h). Strikingly, the sequence of responses is reversed when we set $\alpha = 2$, now reaching black significantly before impacting grey (Fig.\ \ref{Fig1}i).

This diversity of propagation patterns is also expressed by the time-scales of the traveling signal, ranging from $t \sim 10^{-2}$ in case $\alpha = 2$ (green), to $t \sim 1$ for $\alpha = 1$ (blue), reaching $t \sim 10^3$ for $\alpha = 1/2$ (red), several orders of magnitude difference in time-scales exhibited by the same network. Together, these results clearly show that signal propagation is not determined solely by the network topology, but rather by the intricate interplay between this topology and the system's intrinsic dynamics, with even slight changes (value of $\alpha$) having rather dramatic consequences. This illustrates the challenges in predicting information spread in networked systems, where even the relative response times, {\it e.g.,} which nodes respond first and which later, or the typical time scales of the spread, ranging from $10^{-2}$ to $10^3$, are seemingly diverse and unpredictable.
  
\subsection*{Dynamic classes of propagation}

\noindent
To advance from the specific observation above towards a systematic investigation of network signal propagation, we seek to separate the role of the network topology versus that of the dynamics. Therefore, we constructed a systematic \textit{testing ground} combining a diverse body of model and empirical networks with a set of frequently encountered dynamical models. This includes Erd\H{o}s-R\'{e}nyi (ER) and scale-free \cite{Barabasi1999} networks with different link weight distributions (SF, SF1, SF2), as well as empirical networks from social \cite{Opsahl2009,Eckmann2004}, biological \cite{Yu2008,Rual2005} and ecological \cite{Robertson1927} domains. To scan the dynamics \textit{space}, we collected relevant dynamic models, capturing epidemic spreading ($\E$ \cite{PastorSatorras2001a,Hufnagel2004,Dodds2005}), ecological interactions ($\M$ \cite{Gao2016}), regulatory dynamics ($\R_1$, $\R_2$ \cite{Alon2006,Karlebach2008}) and population dynamics ($\P_1$, $\P_2$ \cite{Gardiner2004,Novozhilov2006,Hayes2004}), together a broad spectrum of nonlinear models from diverse application fields. We arrive at a combination totaling $36$ systems - each pairing a network with its relevant dynamics, \textit{e.g.}, ECO with population dynamics $\P$ - comprising together a rich testing ground on which we can systematically observe and decipher the potential signal propagation patterns (Fig. \ref{Fig2}a,b). 

Introducing activity perturbations, as in Fig.\ \ref{Fig1}, we examined signal propagation in each of our $36$ combined networks/dynamics. An example of the results, obtained from the weighted scale-free network SF, across all six dynamic models, is presented in Fig.\ \ref{Fig3}a - f. As before, we find that despite the fact that the networks and layouts in all panels are identical, the spatio-temporal propagation patterns are visibly different, depending on the type of dynamics: in some cases propagating from the core to the periphery ($\R_1, \P_1$, blue), in others advancing from the periphery inwards ($\R_2, \P_2$, red) and finally, in $\M$ and $\E$, featuring a seemingly random scatter of early responding nodes (green).  

To quantitatively analyze these different spreading patterns we measure the propagation time $\T ij$ for the signal in $j$ to reach the target node $i$. This is captured by $\Delta x_i(t = \T ij) = \eta \Delta x_i(t \rightarrow \infty)$, namely $\T ij$ represents the time when $i$ has reached an $\eta$-fraction of its final response to the $j$-signal (typically setting $\eta \sim 1/2$, the {\it half-life} of $i$'s response; Fig.\ \ref{Fig1}b and Supplementary Section 3.2). We then measured the probability density function $P(T)$ for $\T ij$ to be between $T$ and $T + \dif T$. In Fig.\ \ref{Fig3}g - l we show the resulting density functions as obtained from the ER (top) and SF (bottom) networks. We find that the diversity of observed propagation patterns condenses around three highly distinctive classes of spatio-temporal spread, helping us systematically categorize the observed \textit{zoo} of propagation patterns:

\noindent
\textbf{Distance driven propagation} (Fig.\ \ref{Fig3}g,h, blue).
For $\R_1$ and $\P_1$ the density $P(T)$ is identical in both ER (top) and SF (bottom), indicating that $\T ij$ is unaffected by the network's degree distribution. The sharp peaks in $P(T)$ express the fact that the propagation occurs in discrete time intervals, corresponding to the countable steps along the paths between node pairs. Hence the spatio-temporal propagation is driven by the path length $\m Lij$ between the source and the target nodes. Indeed, Fig.\ \ref{Fig3}m,n shows that $\T ij$ is linearly dependent on $\m Lij$, confirming the distance driven propagation. Such dynamics, in which the propagation is naturally depicted by the network paths, has been previously observed, \textit{e.g.}, in disease propagation \cite{Brockmann2013}, yet, our results expose that it represents but one of a variety of potential propagation patterns.  

\noindent
\textbf{Degree driven propagation} (Fig.\ \ref{Fig3}i,j, red).
$\R_2$ and $\P_2$ portray a fundamentally different propagation pattern, with $P(T)$ unaffected by the discrete nature of $\m Lij$. The weighted degree distribution, on the other hand, has a profound effect on $P(T)$: for ER (top) we find that $P(T)$ is bounded, while for SF (bottom) it is extremely heterogeneous, with $\T ij$ spanning several orders of magnitude. This represents a degree-driven  propagation, in which the weighted degree distribution is the main determinant of the spatio-temporal spreading patterns. Consequently, we find that $\T ij$ is almost independent of $\m Lij$, and in fact, for the SF network, even decreases with distance - a striking disparity between the network topology and the actual patterns of information transfer (Fig.\ \ref{Fig3}o,p). 

\noindent
\textbf{Composite propagation} (Fig.\ \ref{Fig3}k,l, green).
The third class is represented by $\M$ and $\E$, where $P(T)$ is affected both by $\m Lij$ and by the SF topology. To observe this we show both $P(T)$, the general $\T ij$ distribution, and $P(T|\m Lij)$, which represents $\T ij$ at given distances, depicted by the inner peaks in different shades of green. We find that $P(T|\m Lij)$ shows a distance driven delay, with the peak density successively progressing as $\m Lij$ is increased. On the other hand, the variance of these inner peaks depends on the degrees, narrow in ER, and broad, and therefore overlapping, in SF. Hence, the resulting spreading patterns are a composition of network distance and degree heterogeneity: on the one hand $\T ij \sim \m Lij$, as confirmed by Fig.\ \ref{Fig3}q,r, a distance driven feature, yet on the other hand, within each $\m Lij$ shell we observe heterogeneity (\textit{i.e.} variance of $P(T|\m Lij)$), that is driven by the bounded/fat-tailed nature of the degree distribution.

This classification represents our first key observation, advancing us towards systematically understanding the rules of information propagation on networks. It indicates that the irregular and seemingly unpredictable propagation presented in Figs.\ \ref{Fig1}d - f and \ref{Fig3}a - f, features recurring characteristic patterns, suggesting the existence of hidden rules that bind together these diverse behaviors. While Fig.\ \ref{Fig3} covers our $24$ model systems, in Supplementary Section 4 we further verify these dynamics classes on our set of $12$ additional empirical systems, detailed in Fig.\ \ref{Fig2}a.

Along the way our classification exposes a delicate balance between diversity and universality, whose theoretical roots we explore below: (i) identical networks may exhibit highly distinctive spreading patterns, depending on the dynamics; (ii) different networks (SF vs.\ ER) may sometimes follow similar propagation patterns (Fig.\ \ref{Fig3}g,h); (iii) the observed propagation patterns can be binned into discrete universality classes (blue, red, green), with similar behavior within each class. Next, we show that this extremely rich behavior can be analytically derived from the complex interplay between the network structure and the system's intrinsic nonlinear interaction dynamics. 

\subsection*{Analytically predicting the patterns of spatio-temporal propagation}

\begin{figure}[t]
\centering
\includegraphics[angle=0,width = 17cm]{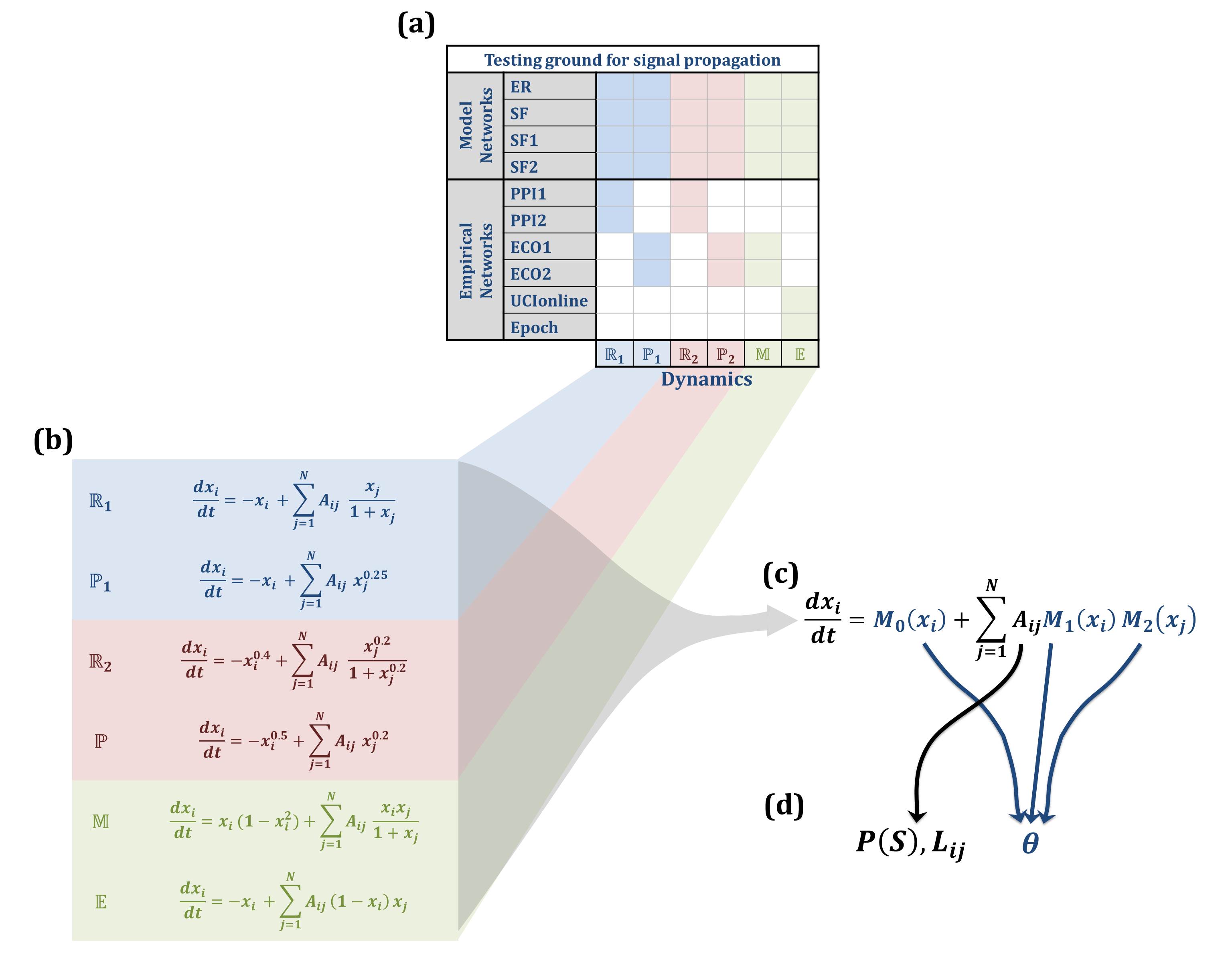}
\caption{
{\bf Testing ground for network signal propagation}. 
(a) 
We tested signal propagation on $36$ relevant combinations constructed from $10$ networks and $6$ dynamic models, for example, epidemic spreading $\E$ on the social networks  UCIonline and Epoch (shaded boxes). The networks (Supplementary Section 3.4): ER - Erd\H{o}s-R\'{e}nyi; SF, SF1, SF2 - scale-free networks with binary, uniformly and scale-free distributed weights, respectively; PPI1/2 - yeast/human protein interaction networks \cite{Yu2008,Rual2005}; ECO1/2 - plant pollinator network of Carlinville Illinois \cite{Robertson1927}, collapsed on to the plants/pollinators; UCIonline - University of California Irvine online instant messaging network \citep{Opsahl2009}; Epoch - email correspondence network \cite{Eckmann2004}. We tested all dynamic models on our four model networks ($24$ shaded boxes - top) and on the appropriate empirical networks ($12$ shaded boxes - bottom).  
(b)
The dynamics (Supplementary Section 2): $\R_1, \R_2$ represent gene regulation via the Michaelis-Menten model \cite{Alon2006,Karlebach2008} with different exponents for the self-dynamics ($1$ vs.\ $0.4$) and for the regulating Hill function ($1$ vs.\ $0.2$); $\P_1, \P_2$ capture population dynamics through birth-death processes \cite{Novozhilov2006,Gardiner2004,Hayes2004}; $\M$ describes mutualsitic interactions, \textit{e.g.}, plant-pollinator relationships in ecological networks \cite{Gao2016} and $\E$ is the susceptible-infected-susceptible (SIS) model for epidemic spreading \cite{Hufnagel2004,PastorSatorras2001a,Dodds2005}.
(c)
We offer to capture all these dynamics, as well as a broad family of additional pairwise dynamics \cite{Barzel2013} through the universal equation (\ref{Dynamics}). Its generic nonlinear terms capture the mechanisms driving each node's self dynamics ($M_0(x)$) and its pairwise interaction with its direct neighbors ($M_1(x), M_2(x)$).
(d)
The propagation patterns emerge from the interplay of the weighted topology $\m Aij$ and the system's intrinsic dynamics $\mathbf{M} = (M_0(x), M_1(x), M_2(x))$. The topology provides the path lengths $\m Lij$ and the weighted degree distribution $P(S)$; the dynamics determine how these topological features translate into $\tau_i$ through $\theta$ (\ref{Taui}). Combining the two contributions, \textit{e.g.}, Eq.\ (\ref{Lij}), provides the spatio-temporal propagation $T(j \rightarrow i)$.   
}
\label{Fig2}
\end{figure}

\begin{figure}[t]
\centering
\includegraphics[angle=0,width = 14cm]{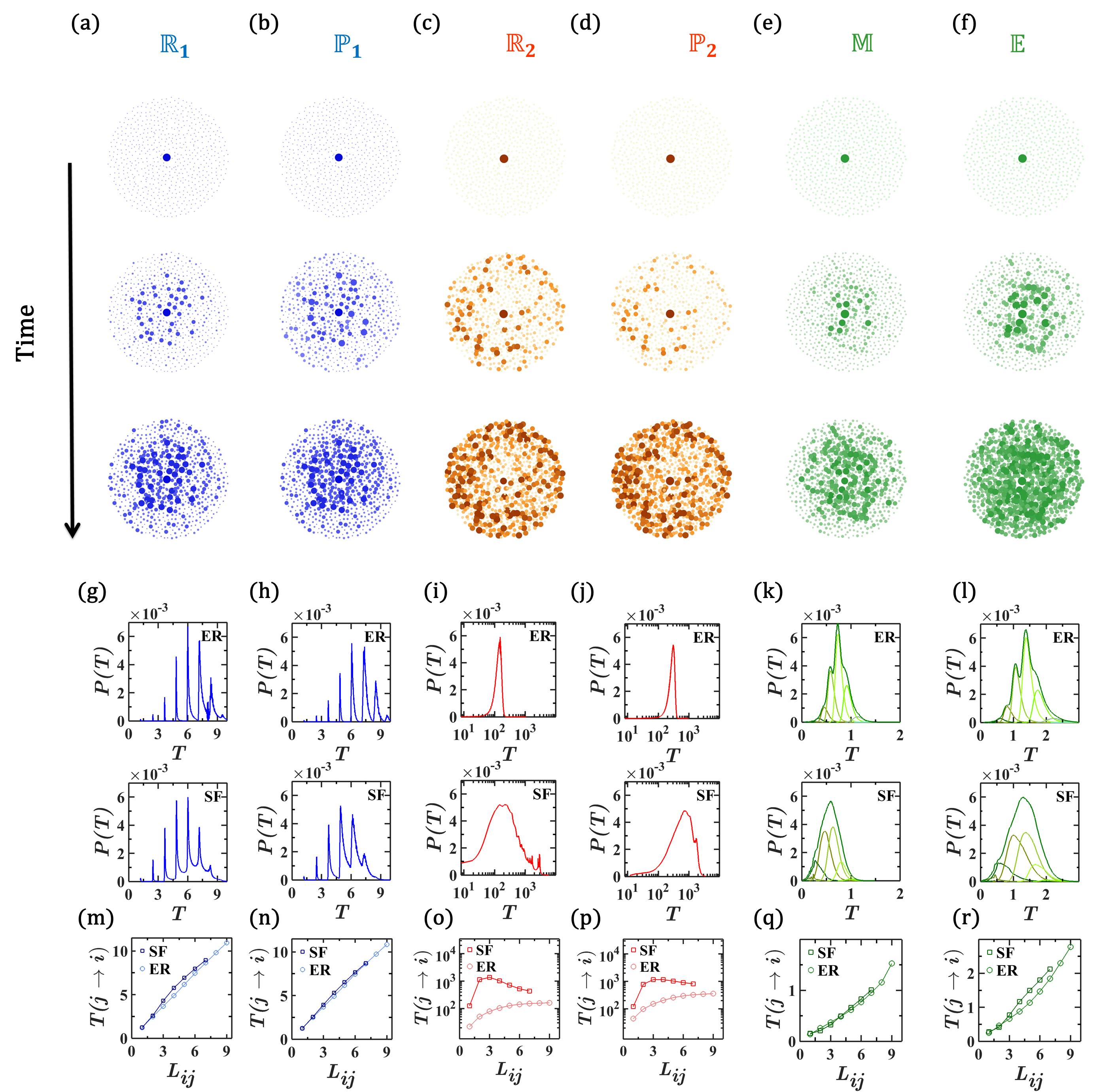}
\vspace{-2mm}
\caption{
\textbf{Classifying the \textit{zoo} of propagation patterns}.
(a)
Propagation on the weighted scale-free network SF under regulatory dynamics $\R_1$. At $t = 0$ we introduce a perturbation in the activity of a randomly selected source node (center), then track the propagation, presenting three snapshots observed at selected time points. The size and color depth of each node represent its response, hence nodes that received the signal at earlier times appear first.    
(b) - (f)
We repeated this experiment on the same network and the same source node, using different dynamic models (Fig.\ \ref{Fig2}b). We observe different propagation patterns depending on the dynamics, resulting in a \textit{zoo} of seemingly unpredictable propagation patterns, reminiscent of the ones observed in the specific example of Fig.\ \ref{Fig1}d - f.
(g) - (h)
The probability density function $P(T)$ vs.\ $T$ as obtained from $\R_1$ and $\P_1$ on ER (top) and SF (bottom). We find that $P(T)$ exhibits multiple sharp peaks in both ER and SF, indicating that the (weighted) degree distribution $P(S)$ has little impact on the propagation.
(i) - (j)
In $\R_2$ and $\P_2$ $P(T)$ has a fundamentally different form with no discrete peaks. Here $P(T)$ is broader in SF (bottom) compared to ER (top), showing that now $P(S)$ has a significant impact on the propagation.
(k) - (l)
$\M$ and $\E$ exhibit a third class with $P(T)$ featuring multiple overlapping peaks. To highlight these peaks we show (in shades of green) $P(T| \m Lij)$, capturing the distribution of $\T ij$ among $i,j$ pairs at equal distance, {\it i.e.} all pairs at distance $\m Lij = 1,2,\dots$. The total density $P(T)$ equals to the sum of these peaks. The inner peaks are broader and hence overlap in SF (bottom) compared to ER (top). Therefore in this class the propagation is affected both by distance, \textit{i.e.} discrete peaks, and by $P(S)$, \textit{i.e.} variance within each peak. 
(m) - (n)
$\T ij$ vs.\ $\m Lij$ exhibits a linear relationship for both SF and ER.
(o) - (p)
$\T ij$ is almost independent of $\m Lij$ in the case of $\R_2$ and $\P_2$, exhibiting a propagation that is indifferent to network distance.
(q) - (r)
For $\M$ and $\E$ we again have $\T ij \sim \m Lij$. Together our analysis shows that the diverse propagation patterns of (a) - (f) categorize into three discrete classes - blue (distance driven), red (degree-driven) and green (composite). Similar results from all our $36$ model/empirical systems, \textit{i.e.} the testing ground of Fig.\ \ref{Fig2}, appear in Supplementary Section 4.
}
\label{Fig3}
\end{figure}

\noindent
To understand the roots of the observed propagation patterns we develop a general formulation, that can capture, within a unified framework, the behavior of all the diverse dynamic models used in Figs.\ \ref{Fig1} and \ref{Fig3}. Therefore we consider the universal equation (Fig.\ \ref{Fig2}c)
\vspace{1mm}
\begin{equation}
\dod{x_i}{t} = 
M_0 \big( x_i(t) \big) + 
\sum_{j = 1}^{N} \m Aij M_1 \big( x_i(t) \big) M_2 \big( x_j(t) \big),
\label{Dynamics}
\end{equation}
\noindent
in which the nonlinear functions $\mathbf{M} = (M_0(x), M_1(x), M_2(x))$ can cover each of the systems included in our testing ground (Fig.\ \ref{Fig2}a,b), as well as a broad range of additional steady-state dynamics, in the context of social \cite{Castellano2009}, biological \cite{Voit2000,May1976b}, neuronal \cite{Stern2014,Li2016} and technological \cite{Hayes2004} interactions. For instance, the regulatory models $\R_1, \R_2$ are covered by (\ref{Dynamics}) through $\mathbf{M} = (-Bx^{\alpha}, 1, x^h/(1 + x^h))$; similarly, the classic susceptible-infected-susceptible (SIS) model ($\E$) can be cast into (\ref{Dynamics}) using $\mathbf{M} = (-Bx, 1 - x, x)$. Therefore Eq.\ (\ref{Dynamics}) provides a universal description of network dynamics, applicable for a broad range of relevant systems.

To link the dynamics (\ref{Dynamics}) to the observed spatio-temporal propagation patterns, we first focus on each node's individual response time $\tau_i$ to a directly incoming signal. Indeed, the signal propagation time, $\T ij$, which captures the complete spatio-temporal propagation, is an aggregation of all individual responses along the trajectory from $j$ to $i$. Hence predicting $\tau_i$ can help us construct the desired $\T ij$ as a sequence of individual responses. In Supplementary Section 1 we show, based on linear response theory, that we can link $\tau_i$ to $i$'s weighted degree $S_i = \sum_{j = 1}^N \m Aij$ through the universal scaling relationship 
%
%
\begin{equation}
\tau_i \sim S_i^{\theta},
\label{Taui}
\end{equation}
%
%
\noindent
where 
%
%
\begin{equation}
\theta = -2 - \Gamma(0).
\label{Theta}
\end{equation}
\noindent
The parameter $\Gamma(0)$ is fully determined by the system's dynamics $\mathbf{M} = (M_0(x), M_1(x), M_2(x))$ through the leading powers of the Hahn series expansion \cite{Hahn1995}
\vspace{1mm}
\begin{equation}
Y \left( R^{-1}(x) \right) = 
\sum_{n = 0}^{\infty} C_n x^{\Gamma(n)},
\label{ZRminus1}
\end{equation} 
%
%
\noindent
where $Y(x) = \left( \dod{[M_1R]}{x} \right)^{-1}$, $R(x) = -M_1(x)/M_0(x)$ and $R^{-1}(x)$ denotes its inverse function. The Hahn expansion in (\ref{ZRminus1}) is a generalization of the Taylor expansion, to include both negative and real powers; hence $\Gamma(n)$, $n = 0,\dots,\infty$, represents a sequence of real powers in ascending order, \textit{i.e.} $\Gamma(n + 1) > \Gamma(n)$. Equation (\ref{Theta}) relates the exponent $\theta$ in (\ref{Taui}) to the leading power $\Gamma(0)$ of (\ref{ZRminus1}), hence directly linking $\tau_i$ to the system's dynamics $\mathbf{M}$ (see Supplementary Section 2 for detailed application of (\ref{Theta}) and (\ref{ZRminus1}) on all dynamics of Fig.\ \ref{Fig2}b).   

\begin{figure}[t]
\centering
\includegraphics[angle=0,width = 9.5cm]{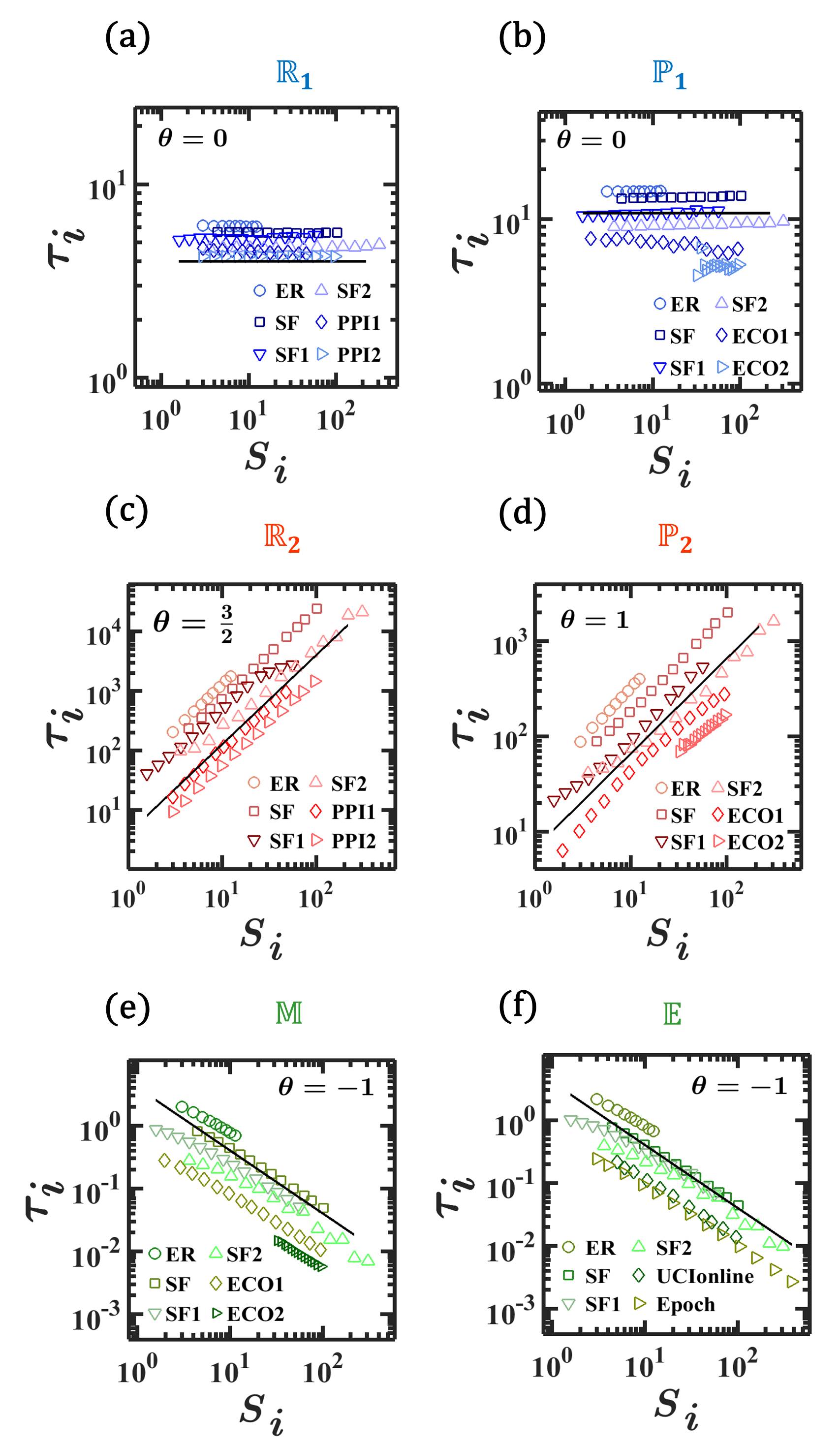}
\vspace{-2mm}
\caption{
{\bf Dynamic universality classes of signal propagation}.
We measured the local response times $\tau_i$ of all nodes vs.\ their weighted degree $S_i$, for our $36$ networks and dynamics, detailed in Fig.\ \ref{Fig2}a.
(a) - (b)
For $\R_1$ and $\P_1$ (symbols) we find that $\tau_i \sim S_i^{\theta}$ with $\theta = 0$ (black solid line), in perfect agreement with the prediction of Eqs.\ (\ref{Taui}) and (\ref{Theta}). This scaling relationship is sustained across diverse model (ER, SF, SF1, SF2) and empirical networks (PPI1, PPI2, ECO1, ECO2), confirming that $\theta$ is independent of $\m Aij$. 
(c) - (d)
For $\R_2$ and $\P_2$ we predict $\theta = 3/2$ and $\theta = 1$ respectively (solid lines), in perfect agreement with the observed results (symbols).
(e) - (f)
For $\M$ and $\E$ we predict $\theta = -1$, as confirmed for both the model and relevant empirical networks. The value of $\theta$ defines the dynamic universality class of each system, determined by the dynamics through (\ref{Theta}), and independent of $\m Aij$, hence grouping together highly distinct networks, that feature the exact same scaling relationship within each dynamic class (panels). This scaling relationship helps us bridge between the topological characteristics $S_i, P(S)$ and their dynamic translation into $\tau_i, P(T)$, and ultimately $\T ij$ (Eq.\ \ref{Lij})). Data points represent logarithmic bins \cite{Milojevic2010} in $S_i$ (Supplementary Section 3.3).
}
\label{Fig4}
\end{figure}

Equations (\ref{Taui}) - (\ref{ZRminus1}) represent our first analytical prediction, showing that the individual response times of all nodes are driven by the interplay between the topology $\m Aij$, through $S_i$ in (\ref{Taui}), and the dynamics $\mathbf{M}$ through $\theta$ (\ref{Theta}). Therefore, the exponent $\theta$ advances us towards our main goal: it helps us translate the static network structure into dynamic insight, by mapping a node's temporal response ($\tau_i$), a dynamic property, to that node's weighted degree ($S_i$), a topological characteristic. To test this prediction, we measured $\tau_i$ vs.\ $S_i$ for each of the $36$ systems summarized in Fig.\ \ref{Fig2}a. The results, presented in Fig.\ \ref{Fig4}, are in excellent agreement with our theoretically predicted scaling: for $\R_1$ and $\P_1$ Eq.\ (\ref{Theta}) predicts $\theta = 0$ (Fig.\ \ref{Fig4}a,b); for $\R_2$ and $\P_2$ it predicts $\theta = 3/2$ (Fig.\ \ref{Fig4}c) and $\theta = 1$  (Fig.\ \ref{Fig4}d), respectively; for $\M$ and $\E$ the prediction is $\theta = -1$ (Fig.\ \ref{Fig4}e,f), all perfectly confirmed by our simulation results. 

Another important aspect of our prediction is that $\theta$ is intrinsic to the system's dynamics $\mathbf{M}$, independent of the network topology $A_{ij}$. Indeed, we observe that Fig.\ \ref{Fig4} groups together our $36$ systems into six classes, each exhibiting the exact same scaling relationship (\ref{Taui}), based on their shared dynamics. This exposes a striking universality sustained across diverse networks, ranging in size, density and structural heterogeneity. More broadly, it indicates that $\theta$ is a \textit{fingerprint} of the system's dynamic model, providing the desired separation of topology vs.\ dynamics: the topology ($\m Aij$) determines the degrees $S_i$ and hence the weighted degree distribution $P(S)$; the dynamic model ($\mathbf{M}$) translates these into $\tau_i$ through $\theta$ (Fig.\ \ref{Fig2}d).

Next, we show that $\theta$ (\ref{Theta}) not only provides the local response times $\tau_i$, but also exposes the origins of the three universality classes observed in Fig.\ \ref{Fig3}:

\noindent
\textbf{Distance driven propagation} ($\R_1$, $\P_1$, blue). 
In case $\theta = 0$ we have $\tau_i$ in (\ref{Taui}) independent of $S_i$. This implies that regardless of $P(S)$, fat-tailed or bounded, all nodes exhibit approximately uniform response times. Therefore, as the signal propagates along the network paths, each node in its trajectory causes, on average, the same delay, and hence the propagation time $\T ij$ is primarily governed by the number of nodes along the path from $j$ to $i$, precisely the distance driven propagation observed in Fig.\ \ref{Fig3}m,n. This form of propagation condenses all nodes into discrete {\it shells}, comprising the nearest neighbors of the signal, the next nearest neighbors and so on. In each of these shells, the signal reaches all nodes approximately simultaneously, resulting in the discrete time intervals, which shape the separated peaks of $P(T)$ (Fig.\ \ref{Fig3}g,h). Finally, with response time being independent of degree the structure of $P(T)$ is unaffected by the degree distribution, explaining the similar propagation patterns observed across the highly distinct ER and SF networks.

\noindent
\textbf{Degree driven propagation} ($\R_2$, $\P_2$, red). 
In case $\theta > 0$ Eq.\ (\ref{Taui}) predicts that hubs respond at a slower rate than low degree nodes, in effect being the {\it bottlenecks} of signal propagation. This gives rise to the degree driven propagation observed in Fig.\ \ref{Fig3}i,j where SF networks (bottom) exhibit a much broader $P(T)$, in comparison to ER networks (top), a consequence of the delayed propagation caused by the hubs. The greater is $\theta$, the more pronounced is the effect. In this class the path length $\m Lij$ between the source and the target is of little importance compared to the degrees of the nodes along these paths. Indeed, in SF networks paths are extremely short (of order $\sim \log N$ or smaller \cite{Cohen2003}), while degrees range over orders of magnitude. Consequently, the propagation patterns are dominated by $P(S)$ rather than by $\m Lij$, as confirmed by Fig.\ \ref{Fig3}o,p.

\noindent 
\textbf{Composite propagation} ($\E$, $\M$, green). 
For $\theta < 0$ the hubs respond rapidly, hence signal propagation is primarily limited by the path length from the source to the target. However, within each shell around the signal source we observe a diversity in $\T ij$, driven by the degree heterogeneity ($P(S)$), with hubs responding earlier than small nodes. The result is composite dynamics, combining separated peaks, which overlap due to degree heterogeneity (Fig.\ \ref{Fig3}k,l).

\begin{figure}[t]
\vspace{1.5cm}
\centering
\includegraphics[angle=0,width = 16cm]{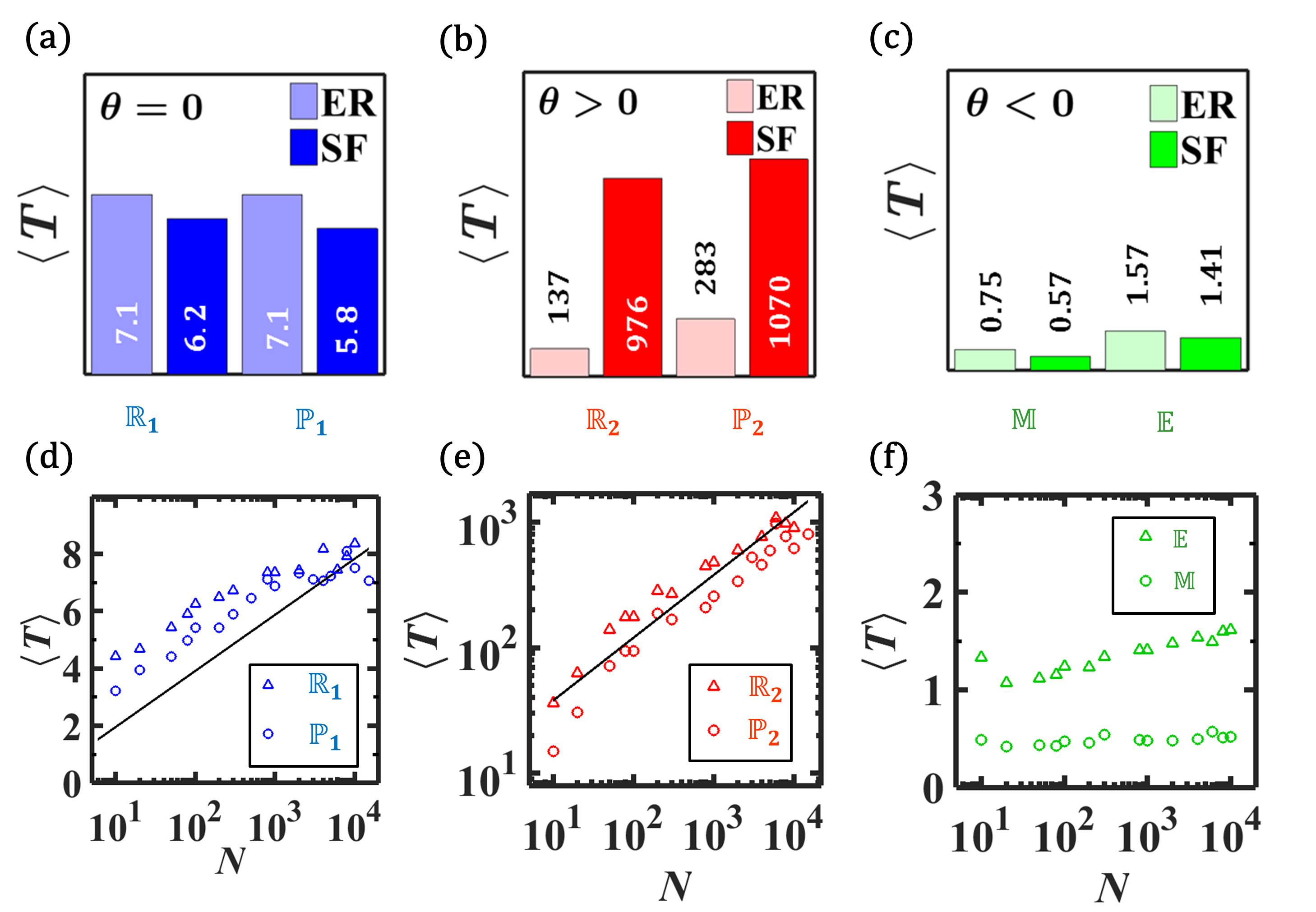}
\vspace{-2mm}
\caption{{\bf The efficiency of signal propagation}.
The average propagation time $\av T$ on an Erd\H{o}s-R\'{e}nyi (ER) and a scale-free (SF) network with identical average degree. 
(a)
For distance driven dynamics ($\R_1$, $\P_1$, $\theta = 0$) $\av T$
is not significantly affected by the ER/SF networks, other than a minor decrease in $\av T$ for SF, a consequence of the typically shorter paths characterizing SF networks \cite{Cohen2003}.
(b)
For degree driven dynamics ($\R_2$, $\P_2$, $\theta > 0$) the hubs delay the propagation, and hence the SF topology translates to a significant increase in $\av T$. Therefore, in this class, degree heterogeneity causes inefficient propagation, slowing the rate of information spread. The effect is more pronounced when $\theta$ is large: indeed, for $\R_2$ ($\theta = 3/2$) we observe a $612\%$ increase in $\av T$, while for $\P_2$ ($\theta = 1$) the delay is less than half, $278\%$. 
(c)
For composite dynamics ($\M$, $\E$, $\theta < 0$) $\av T$ is again unaffected by hubs, dominated mainly by the response time of the small nodes, which is roughly the same in ER and SF. For both ER and SF, however, $\av T$ is much smaller than in the two other classes (blue, red) due to the fast response of the hubs along the pathways from source to target, leading to an ultra-efficient propagation.
(d)
The average propagation time $\av T$ vs.\ the number of nodes $N$ as obtained from $\R_1$ (triangles) and $\P_1$ (circles). In distance driven propagation we find that $\av T \sim \log N$ (solid line), a logarithmic dependence on system size, corresponding to the \textit{efficient spread} predicted by our theory. 
(e)
In degree driven dynamics we predict \textit{slow spread}, in which $\av T \sim N^{\alpha}$ (solid line represents $\alpha = 1/2$). Here, despite the fact that SF shrinks the topological distance $\av L$, it dramatically inflates the temporal distance $\av T$.
(f)
In composite dynamics we predict \textit{ultra-efficient spread}, namely $\av T \sim \rm{const}$, independent of system size. Here $N$ spans four orders of magnitude, while $\av T$ is practically constant, confirming our prediction. Data points represent logarithmic bins \cite{Milojevic2010} in $N$ (Supplementary Section 3.3). 
}
\label{Fig5}
\end{figure}

Hence we find that the {\it zoo} of diverse spreading behaviors observed in Figs.\ \ref{Fig1} and \ref{Fig3} is, in fact, a consequence of a deep universality that can be fully predicted by our formalism through the single, analytically tractable, universal exponent $\theta$ in (\ref{Theta}). This exponent helps shed light on the link between structure and dynamics, a central theoretical challenge in the study of complex systems \cite{Barabasi2002,Barzel2009}. For example, we can now uncover the dynamic consequences associated with two of the most profound characteristics of real networks: (i) most real networks exhibit extremely short paths between all nodes, with the average path length often following $\av {\m Lij} \sim \log N$ \cite{Cohen2003}; (ii) the (weighted) degree distribution $P(S)$ of many real systems is fat-tailed, often scale-free, with highly connected hubs coexisting alongside a majority of low degree nodes \cite{Caldarelli2007}. Here we show that these two topological hallmarks impact the propagation of signals in a rather distinctive fashion. While the short paths accelerate the propagation of signals, the impact of degree heterogeneity depends on the dynamics of the system through $\theta$: hubs may either expedite the propagation of signals ($\theta < 0$, green, Fig.\ \ref{Fig5}c), have no effect on the propagation ($\theta = 0$, blue, Fig.\ \ref{Fig5}a) or cause delays ($\theta > 0$, red, Fig.\ \ref{Fig5}b).
  
To observe the consequences of this interplay between $P(S)$ and $\T ij$, we consider the average propagation time $\av T$, representing the typical time-scale for signals to penetrate the entire network. The smaller is $\av T$ the more efficient is the network in spreading local information. Our dynamic universality classes predict three levels of propagation efficiency: 
{\it Efficient spread} ($\theta = 0, \R_1, \P_1$, Fig.\ \ref{Fig5}d). For distance driven dynamics we have $\T ij \sim \m Lij$, and hence, for a random network $\av T \sim \av {\m Lij} \sim \log N$, a rapid coverage that grows only logarithmically with the system's size. 
{\it Slow spread} ($\theta > 0, \R_2, \P_2$, Fig.\ \ref{Fig5}e). For degree driven dynamics the propagation times are governed by the hubs, whose degrees increase with $N$, hence for a large system ($N \rightarrow \infty$), signals require an extremely long time to penetrate the network. For a scale free network this leads to a scaling behavior $\av T \sim N^{\alpha}$, an inefficient propagation in which $\av T$ diverges polynomially with the size of the system. Therefore, despite the fact that the scale-free property decreases the {\it topological} distance ($\m Lij$) \cite{Cohen2003}, under degree driven dynamics it dramatically \textit{increases} the effective {\it temporal} distance ($\T ij$), emphasizing again the non-trivial translation from topology to dynamics that our theory allows us to predict. 
{\it Ultra-efficient spread} ($\theta < 0, \E, \M$, Fig.\ \ref{Fig5}f). In composite dynamics signals rapidly propagate thanks to the hubs, which effectively shrink the paths between all nodes. Consequently, the propagation time is primarily determined by the response of the target nodes, which is independent of $N$ or of the path length. The resulting propagation is extremely efficient, with $\av T \sim {\rm const}$, being effectively independent of $N$. Indeed, in Fig.\ \ref{Fig5}f we find that networks of vastly different size, ranging over more than four orders of magnitude, are all covered within approximately the same $\av T$, a counter-intuitive form of propagation, that is yet fully predicted by our formalism.

\subsection*{Universal dynamic metric for signal propagation}

\newcommand{\Path}[2]{\Pi (#2 \rightarrow #1) }

\begin{figure}[t]
\centering
\includegraphics[angle=0,width = 14cm]{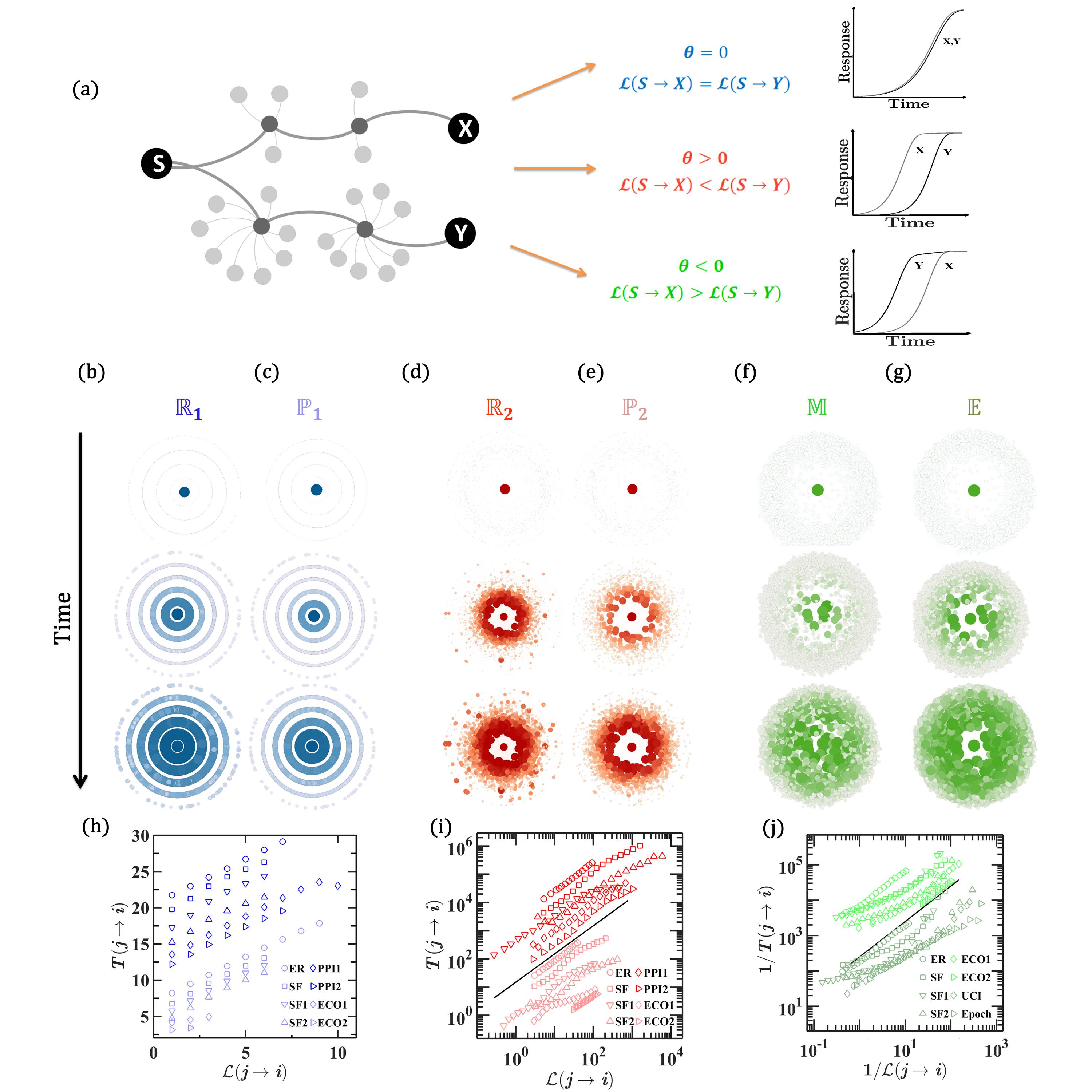}
\vspace{-2mm}
\caption{
{\bf The universal temporal distance $\L ij$}.
(a)
A signal propagating from the source $S$ to two targets $X$ and $Y$, both at distance $\m LSX = \m LSY = 2$. The signal will reach $X$ and $Y$ simultaneously if the dynamics is distance driven (top, blue); $X$ before $Y$ in case of degree driven dynamics (center, red) due to the slowly responding hubs along the path $\Path YS$, and $Y$ before $X$ in composite dynamics (bottom, green). The temporal distance $\L ij$ in (\ref{Lij}) is designed to {\it locate} $X$ and $Y$ at the appropriate distance from $S$, depending on the dynamic class of the propagation.
(b) - (g)
We used $\L ij$ to layout the nodes in each of our $36$ test systems (Fig.\ \ref{Fig2}), here displaying the results obtained for SF, as appear in the original layouts of Fig.\ \ref{Fig3}a - f. The unpredictable and inconsistent behavior, \textit{i.e.} the \textit{zoo} of Fig.\ \ref{Fig3}, transforms into a well-organized concentric propagation pattern, in which the distance from the source naturally captures the actual travel time of the propagating signal. These layouts locate all nodes differentially, according to the system's dynamics. For the distance-driven $\R_1$ and $\P_1$ (blue) nodes are condensed into separated shells, corresponding to the discrete nature of the path lengths (peaks in $P(T)$, Fig.\ \ref{Fig3}). In the degree-driven $\R_2$ and $\P_2$ (red) the hubs become bottlenecks, and hence $\L ij$ assigns a larger distance to paths that are enriched with hubs. In the composite $\M$ and $\E$ our universal $\L ij$ places the slower low degree nodes in the network periphery and shifts hubs towards the center. Additional layouts from our empirical networks are shown in Supplementary Section 4.
(h)
$\T ij$ vs.\ $\L ij$ for all networks under $\R_1$ (dark) and $\P_1$ (light) dynamics. The linear relationship indicates that $\L ij$ precisely captures the actual patterns of propagation. 
(i)
$\T ij$ vs.\ $\L ij$ in the degree-driven $\R_2$ (dark) and $\P_2$ (light). Here, since $\theta > 0$, $\T ij$ and $\L ij$ span several orders of magnitude, hence we use a logarithmic scale (black solid line represents a linear slope). 
(j)
For the composite $\M$ (light) and $\E$ (dark) $\L ij$ and $\T ij$ inversely scale with the weighted degrees of nodes along each path. Therefore we use inverted axes $1/\T ij$ vs.\ $1/\L ij$. In (i) and (j) we employed logarithmic binning \cite{Milojevic2010} (Supplementary Section 3.3). Together (h) - (j) feature results from all our $36$ systems; a specific focus on our $12$ empirical systems appears in Supplementary Section 4.
}
\label{Fig6}
\end{figure}

\noindent
To simplify the observed flow of information we seek a predictive metric, $\L ij$ that transparently reflects the actual propagation times $\T ij$ \cite{Brockmann2013}, namely we seek a temporal distance $\L ij$, for which $\T ij \sim \L ij$. Consider the shortest path from the source $j$ to the target $i$, denoted by the sequence $\Path ij = j \rightarrow p \rightarrow q \rightarrow \dots \rightarrow i$. This path, being shortest, will dominate the spread of the signal $\Delta x_j$ to the target $i$ \cite{Barzel2013}, hence $\T ij$ depends mainly on the travel time along $\Path ij$. We can evaluate this travel time using (\ref{Taui}) to be $\T ij \sim S_p^{\theta} + S_q^{\theta} + \cdots + S_i^{\theta}$, the total lag time accumulated on all nodes along $\Path ij$ (Fig.\ \ref{Fig6}a). In general, we can write

\begin{equation}
\L ij = \min_{\Path ij} \Bigg\{ \sum_{\substack{p \in \Path ij \\ p \ne j}} S_p^{\theta} \Bigg\}
\label{Lij}
\end{equation}
\noindent
where the minimization selects the fastest of all shortest paths between $j$ and $i$. Equation (\ref{Lij}) represents our final result, providing the temporal distance between all pairs of nodes $i$ and $j$, designed to naturally capture the system's dynamic signal propagation. As opposed to other common metrics, $\L ij$ depends not only on the topology, but also on the dynamics $\mathbf M$ through the exponent $\theta$ (\ref{Theta}), therefore accounting for the interplay between structure and dynamics. Hence, for a given $\m Aij$ the distances $\L ij$ are adaptive, relocating all nodes depending on the nature of the system's nonlinear interactions. 

To test (\ref{Lij}) we used it to layout the scale-free networks shown in Fig.\ \ref{Fig3}a - f, placing each node in its appropriate location, at distance $\L ij$ from the perturbed source (Fig. \ref{Fig6}b - g; for the layout of our empirical networks see Supplementary Section 4). The originally unpredictable spreading patterns (Fig.\ \ref{Fig3}) collapse into a concentric propagation, with the desired $\T ij \sim \L ij$. The crucial point is that these layouts, which we predict \textit{a priori}, \textit{i.e.} before observing the simulation results, are dynamically adaptive, appropriately locating the nodes according to the predicted dynamic universality class. Hence, despite using the same $\m Aij$ the nodes are located differently as the dynamics is shifted from  $\R_1$ and $\P_1$ (blue, $\theta = 0$) to $\R_2$ and $\P_2$ (red, $\theta > 0$), and further to $\M$ and $\E$ (green, $\theta < 0$). In Fig.\ \ref{Fig6}h - j we show the observed $\T ij$ vs.\ the analytically calculated $\L ij$ for all of our $36$ model/empirical systems - each in the appropriate class (blue, red, green). We find that (\ref{Lij}) consistently captures the actual patterns of propagation, satisfying the desired $\T ij \sim \L ij$, thus providing a highly predictive, dynamically adaptive universal distance metric for signal propagation (few minor discrepancies appearing in specific systems are discussed in Supplementary Section 4).

\subsection*{Discussion and outlook}

\noindent
Predicting the spread of information in complex networks is at the heart of our ability to understand their dynamic behavior, hence the widespread efforts to collect data on the topology of real biological, social and technological networks. Yet, if we wish to leverage these data into actual dynamic insights, we must systematically translate our findings on network structure into dynamic predictions on information flow. Our formalism offers such translation by separating the contribution of the topology, $\m Aij$, from that of the dynamics, $\mathbf{M}$, through the analytically predicted exponent $\theta$, exposing highly distinctive dynamic universality classes that characterize the connection between $\m Aij$ and the dynamic spreading behavior. The distinctions between these classes are multi-faceted, providing an array of testable predictions, from $P(T)$, through $\tau_i$ (\ref{Taui}) to the scaling of $\av T$ with $N$, highly distinctive features that provide a set of clear observable {\it fingerprints} by which to classify a system's dynamics.

While complex system dynamics can take almost unlimited forms, our formalism shows that the determinants of information spread are restricted to the few leading powers of $\bf M$, as encapsulated within $\Gamma(0)$ in (\ref{ZRminus1}). This groups together fundamentally different dynamics under the same universality class, \textit{e.g.}, ecological interactions ($\M$) and epidemics ($\E$), which exhibit identical spreading patterns - a surprising observation, predicted by (\ref{Theta}).

Most importantly, these powers ($\Gamma(n)$) as opposed to the coefficients ($C_n$) are intrinsic to the system's dynamics, depending on the functional form of $\mathbf{M}$, but not on its specific rate constants. For instance, in the SIS model ($\E$) we have $M_0(x) \sim -x$, $M_1(x) \sim 1 - x$ and $M_2(x) \sim x$ (Fig.\ \ref{Fig2}b). The structure of these three functions, and hence their leading powers, is intrinsic to the dynamic mechanisms of infection and recovery. Therefore our prediction that $\E$ is in the composite universality class ($\theta = -1$, green) is not sensitive to the microscopic {\it rates} of infection/recovery, which vary across different diseases, but rather represents a robust property of the SIS {\it model}, unifying all communicable diseases whose spreading mechanism is captured by the SIS framework. Such universality is a crucial component in our effort to construct a theory of complex system dynamics, as most complex systems are multi-parametric \cite{Gao2016}, allowing no access, or analytical treatment, of their detailed microscopic parameters. Hence we seek empirically observable macroscopic functions that can be directly traced to a small number of the system's relevant and intrinsic parameters, such as the leading powers of $\mathbf{M}$. An analogous approach was successfully employed in the past to expose universality in particle systems \cite{Wilson1975} - we believe that this line of thought may lead to similar breakthroughs in our understanding of complex networked systems.
 
\clearpage

\renewcommand{\bibsection}{\section*{References}}

\bibliographystyle{unsrt}
\bibliography{bibliography}
\clearpage 

\section*{Figures and captions (See full scale images on Page 22)}

\clearpage

\clearpage

\clearpage

\clearpage

\clearpage

\section*{Full scale Figures}

\begin{figure}[b]
\centering
\includegraphics[angle=90,width = 15cm]{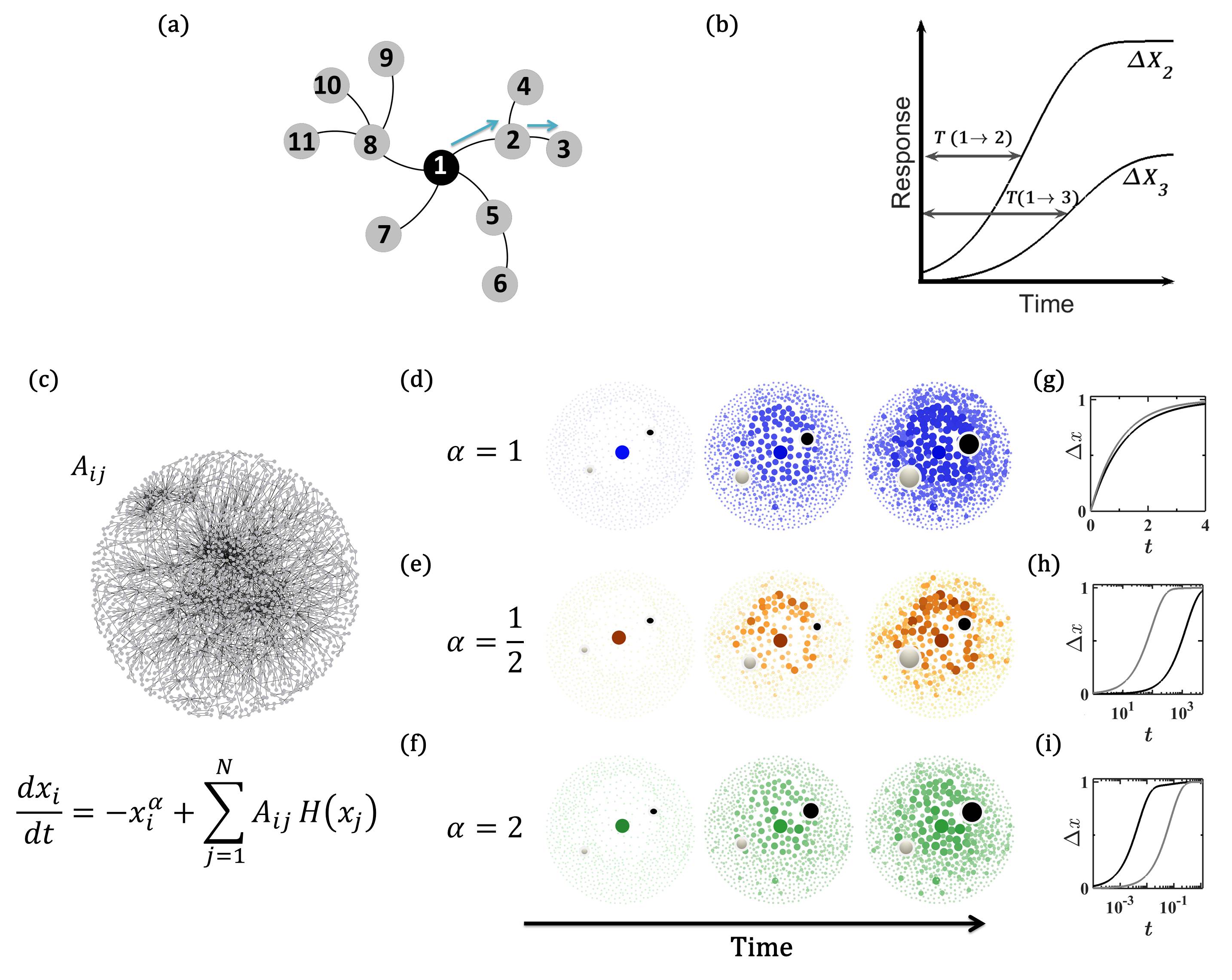}

\vspace{3mm}
{\bf FIG 1. Propagation of signals in a complex networks}.
\end{figure}

\clearpage

\begin{figure}[t]
\vspace{1.5cm}
\centering
\includegraphics[angle=90,width = 15cm]{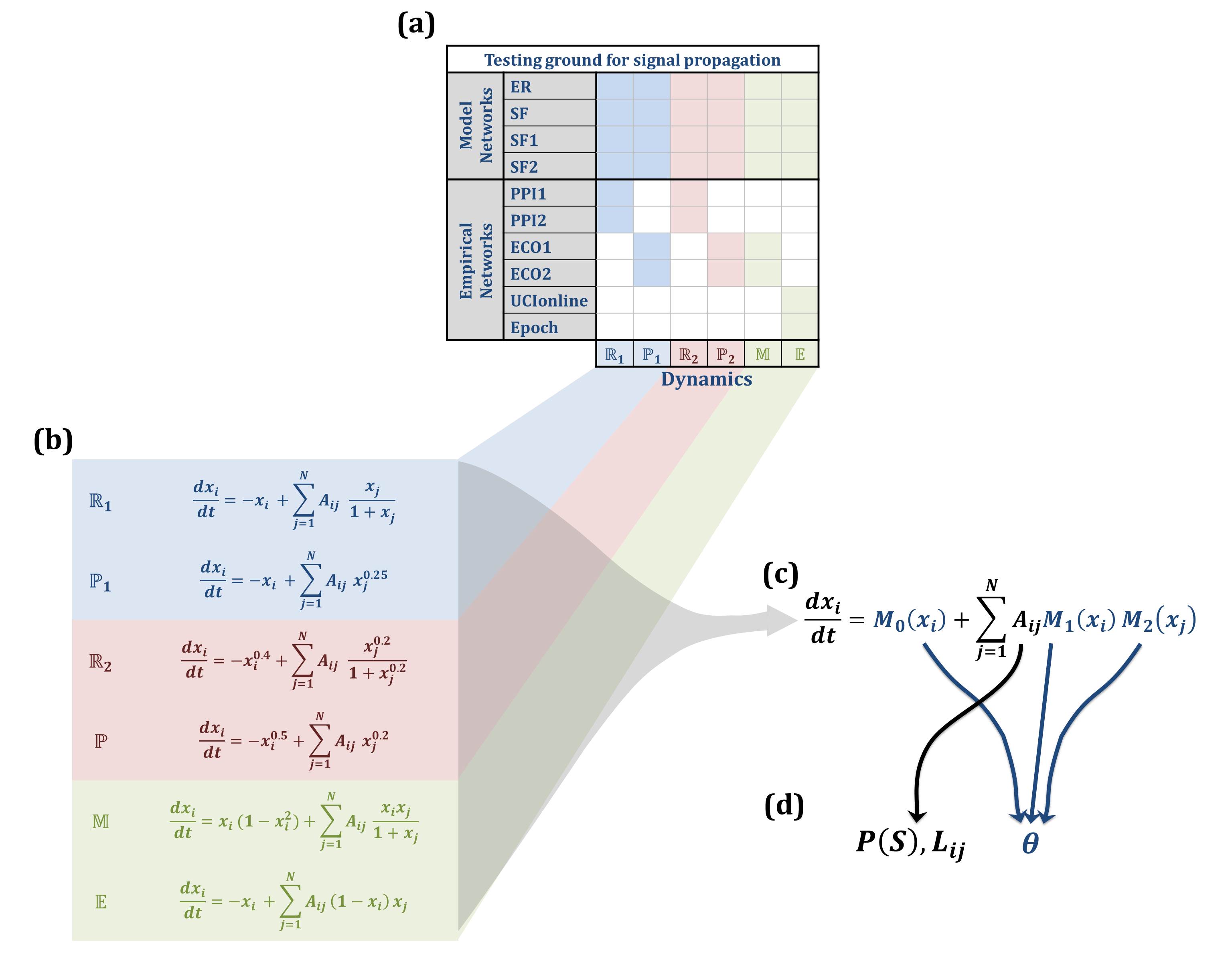}

\vspace{3mm}

{\bf FIG 2. Testing ground for network signal propagation}.
\end{figure}

\clearpage

\begin{figure}[t]
\centering
\includegraphics[angle=0,width = 17cm]{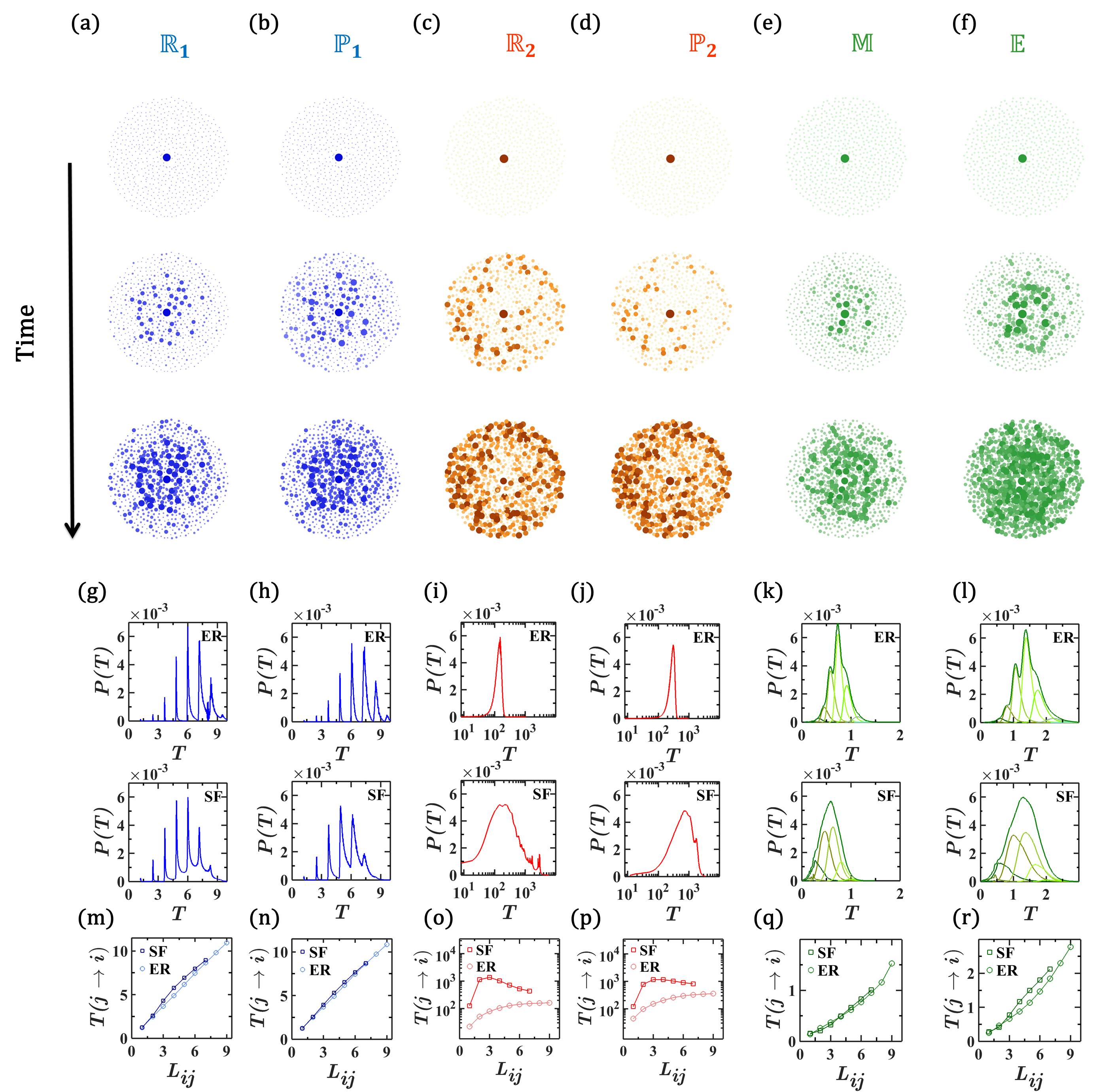}
{

\vspace{3mm}
\textbf{FIG 3. Classifying the \textit{zoo} of propagation patterns}}.
\end{figure}

\clearpage

\begin{figure}[t]
\centering
\includegraphics[angle=0,width = 12cm]{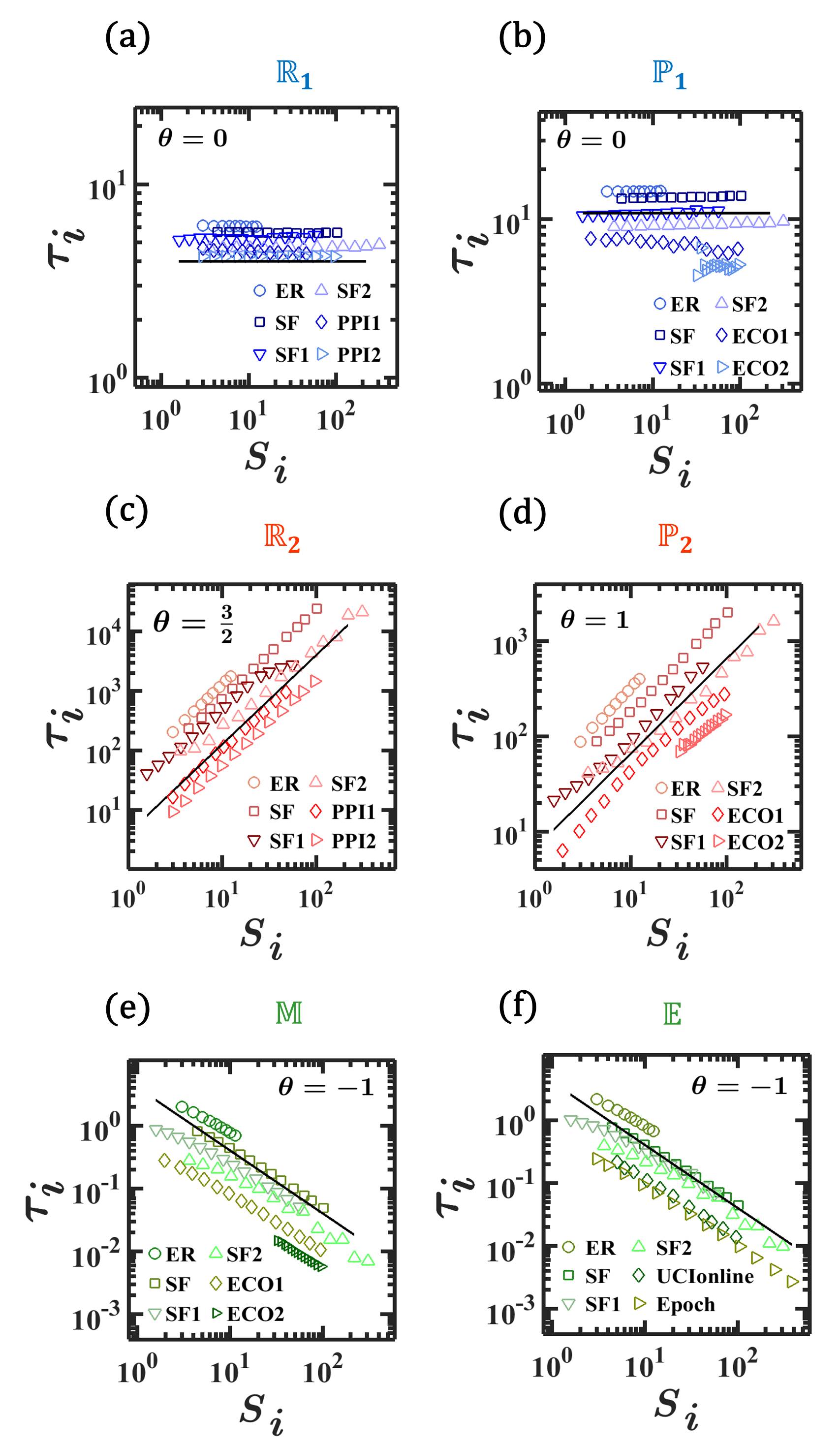}

\vspace{3mm}
{\bf FIG 4. Dynamic universality classes of signal propagation}.
\end{figure}

\clearpage

\begin{figure}[t]
\centering
\includegraphics[angle=90,width = 12cm]{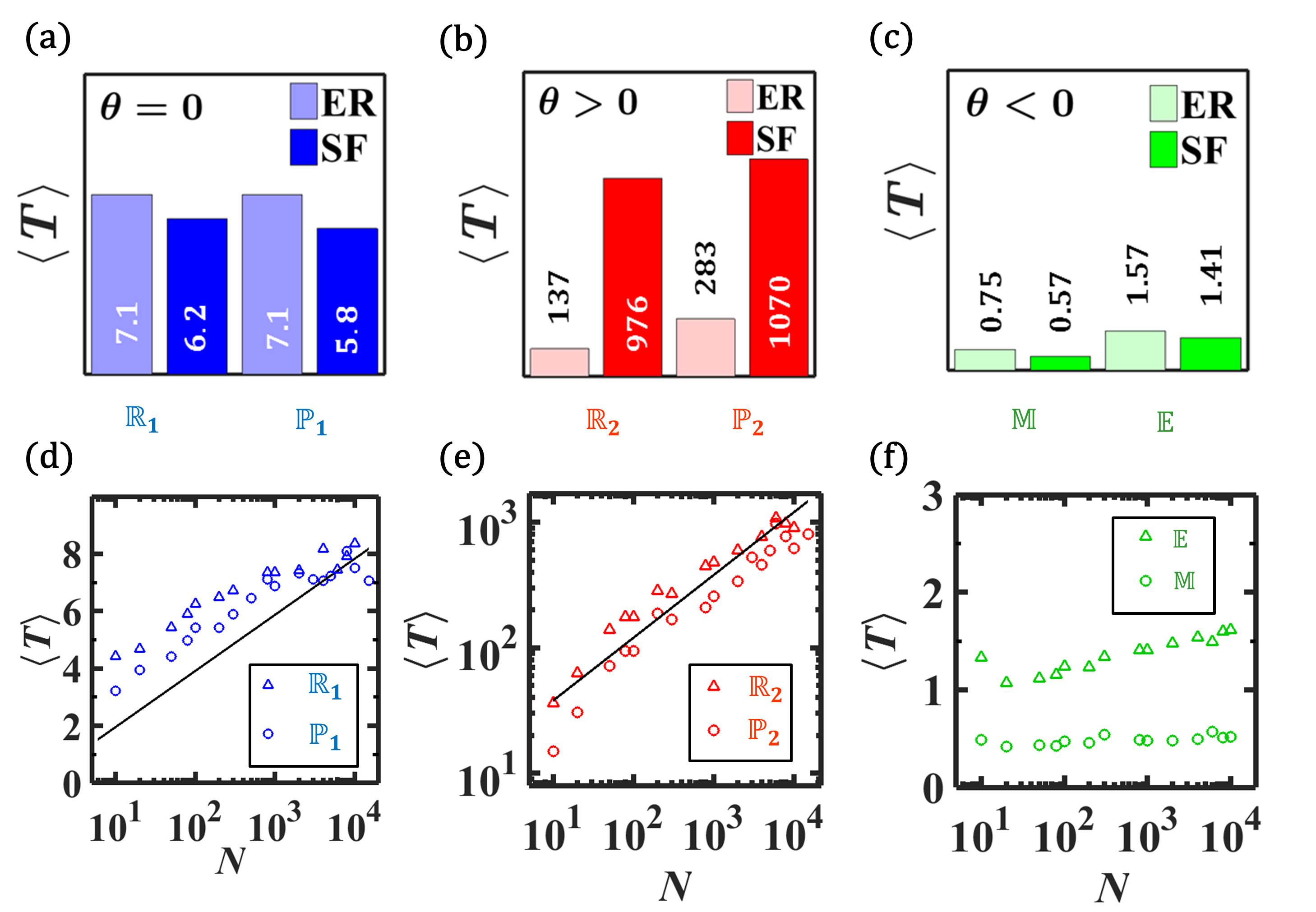}

\vspace{3mm}
{\bf FIG 5. The efficiency of signal propagation}.
\end{figure}

\clearpage

\begin{figure}[t]
\centering
\includegraphics[angle=0,width = 17cm]{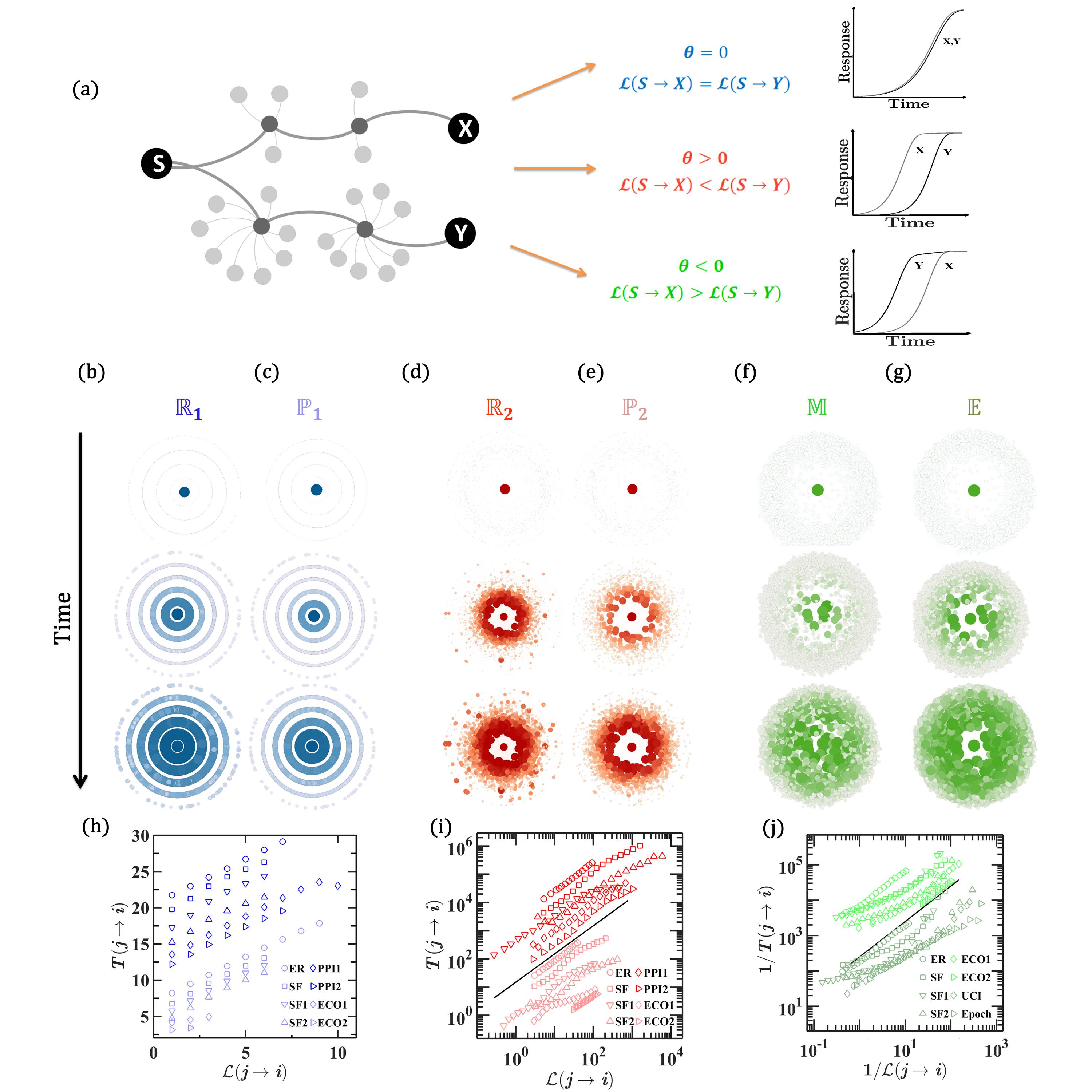}
{

\vspace{3mm}
{\bf FIG 6. The universal temporal distance $\L ij$}.
}
\end{figure}

\end{document}



\renewcommand{\P}
{
\mathbb P
}

\newcommand{\BD}
{
\mathbb P
}

\newcommand{\M}
{
\mathbb M
}

\newcommand{\R}
{
\mathbb R
}

\newcommand{\E}
{
\mathbb E
}

\newcommand{\B}
{
\mathcal B
}

\newcommand{\m}[3]
{
#1_{#2 #3}
}

\newcommand{\av}[1]
{
\left< #1 \right>
}

\newcommand{\D}[2]
{
\mathcal{D}_{#1 #2}
}

\renewcommand{\L}[2]
{
\mathcal{L}{(#2 \rightarrow #1)}
}

\newcommand{\T}[2]
{
T (#2 \rightarrow #1)
}

\newcommand{\G}
{
\mathbf G
}

\newcommand{\Gnn}
{
\mathbf G_{\rm nn}
}

\title{Predicting the patterns of spatio-temporal signal \\ propagation in complex networks\\ Supplementary Material}
\author{Chittaranjan Hens, Uzi Harush, Reuven Cohen \& Baruch Barzel}
\maketitle
\thispagestyle{empty}

\pagebreak
\tableofcontents
\thispagestyle{empty}
\pagebreak
\pagestyle{plain}

\pagenumbering{arabic}

\pagebreak

 \maketitle

\section{Analytical derivations - from $\mathbf{M}$ to $\tau_i$}
\label{Transient}

\renewcommand{\d}{{\rm d}}

To construct the propagation times $\T ij$ we must first obtain the individual response times $\tau_i$, capturing the transient response of each node $i$ to direct incoming perturbations from its interacting neighbors. Indeed, as we show in the main paper (Eq.\ (5)), the propagation times $\T ij$ can be constructed from the sequence of local responses $\tau_i$ along each path, giving rise to the observed universality classes. Hence we use a perturbative approach to derive the response time of a node to a neighboring perturbation $\Delta x_j$. Starting from the dynamic equation

\begin{eqnarray}
\dod {x_{i}}{t} &=& M_{0}(x_{i}(t)) + \sum_{j=1}^{\mathrm{N}} \m Aij M_{1}(x_{i}(t))M_{2}(x_{j}(t)),  
\label{Dynamics}   
\end{eqnarray}

\noindent
we obtain the steady state $x_i$ by setting the derivative on the l.h.s.\ to zero, and then introduce a time-independent perturbation $x_m(t) = x_m + \Delta x_m$ on the activity of node $m$, one of $i$'s nearest neighbors. Node $i$'s response will follow

\begin{equation}
x_i(t) = x_i + \Delta x_i(t),
\label{dxi}
\end{equation}

\noindent 
with $\tau_i$ representing the relaxation time of $\Delta x_i(t)$. Below, we show in detail how to calculate $\tau_i$.

\begin{Frame}
\textbf{Our theoretical framework}. 
Our derivation predicts the scaling relationship between $\tau_i$ and each node's weighted degree $S_i$, directly from the system's dynamics $\mathbf{M} = (M_0(x), M_1(x), M_2(x))$ in (\ref{Dynamics}). Throughout this derivation we rely on two main approximate assumptions:
\begin{itemize}
\item
\textit{Perturbative limit}. We assume that the signal $\Delta x_m$ is small, namely we take the limit $\Delta x_m \rightarrow \dif x_m$, allowing us to employ the tools of linear response theory to treat (\ref{Dynamics}) analytically.
\item
\textit{Configuration model}. We allow $A_{ij}$ to feature any arbitrary degree/weight distribution, including scale-free or other fat-tailed density functions, but assume that it is otherwise random \cite{Newman2010}. Such approximation may overlook certain characteristics pertaining to the network's fine-structure, such as degree-degree correlations \cite{Newman2002}, or clustering, which, in the limit of sparse networks ($\langle k \rangle \ll N \rightarrow \infty$) become negligible due to the random connectivity.   
\end{itemize} 
In Sec.\ \ref{Pertubation and Degree Correlation} we systematically test the robustness of our predictions against these approximations. We examine the impact of large perturbtaions, including the system's response to complete node knockout, an unambiguously \textit{large} perturbation. We also observe our theory's performance under increasing levels of degree-correlations and clustering. We find, that our predictions are highly insensitive to these approximations, successfully withstanding empirically relevant levels of clustering and degree-correlations as well as large signals, all of which have but a marginal - and in fact non-visible - effect on our predicted scaling and universality classes. The origins of this robustness are also discussed in Sec.\ \ref{Pertubation and Degree Correlation}. 

\end{Frame}

\subsection{Configuration model} 

Throughout our analysis below we use the configuration model framework to analyze 
$\m Aij$ \cite{Newman2010}.
Within this framework $\m Aij$ represents a general weighted network with arbitrary degree and weight distributions, but otherwise random structure. 
Hence we assume negligible correlation between the number of neighbors of a node
$k_i$, 
and its link weights
$\m Aij$, 
namely
$P(\m Aij = a | k_i) = P(\m Aij = a)$.
Another significant implication of the configuration model assumption is that we neglect minor structural correlations between nodes and their immediate environment. 
For instance, while two nodes, 
$i$ and $j$,  
may have extremely different topological characteristics, say 
$i$ is a hub and $j$ is a low degree node,
their neighborhoods are assumed to share similar statistical properties, namely 
$i$'s 
(many) neighbors are extracted from the same statistical pool as 
$j$'s 
(few) neighbors.
Specifically, let us denote by 
$\G(S)$
the group of all nodes whose weighted degrees are between
$S$ and $S + \dif S$.
This group can be characterized by one or more random variables
$Q_i$, 
capturing, for instance the activity 
$x_i(t)$ or the relaxation time $\tau_i$ 
associated with a randomly selected node 
$i \in \G(S)$.
The corresponding distribution 

\begin{equation}
P_S(Q_i = q) = P \big(Q_i = q \big| i \in \G(S) \big)
\label{PSQi}
\end{equation}

\noindent
is unique to $\G(S)$, since nodes in $\G(S)$ are distinct from nodes in $\G(S^{\prime})$,
hence, in general
$P_S(Q_i = q) \ne P_{S^\prime}(Q_i = q)$.
This distinction translates also to statistical properties extracted from 
$\G(S)$, for instance the mean value of $Q_i$, expressed by

\begin{equation}
Q(S) = \frac{1}{|\G(S)|}\sum_{i \in \G(S)} Q_i,
\label{Qs}
\end{equation}

\noindent
($|\G(S)|$
represents the number of nodes in 
$\G(S)$)
may differ from 
$Q(S^{\prime})$.
For example, the typical response time of nodes in 
$S$ is potentially different than that of nodes in $S^{\prime}$.

Next we consider the random variable

\begin{equation}
Q_{i,\odot} = \dfrac{1}{S_i} \sum_{n = 1}^N \m Ain Q_n,
\label{Qnn}
\end{equation}

\noindent
a weighted average over $i$'s nearest neighbors, whose probability distribution is given by
$P(Q_{i,\odot} = q)$.
Averaging over nodes in $\G(S)$ we obtain

\begin{equation}
Q_{\odot}(S) = \frac{1}{|\G(S)|}\sum_{i \in \G(S)} Q_{i,\odot} = 
\frac{1}{|\G(S)|}\sum_{i \in \G(S)} \dfrac{1}{S} \sum_{n = 1}^N \m Ain Q_n,
\label{Qnns}
\end{equation}

\noindent
analogous to $Q(S)$ in (\ref{Qs}).
According to the configuration model the nearest neighbors of 
$i \in \G(S)$ and $j \in \G(S^{\prime})$
follow similar statistics, hence we have

\begin{equation}
P \big( Q_{i,\odot} = q \big| i \in \G(S) \big) = 
P \big( Q_{i,\odot} = q \big| i \in \G(S^{\prime}) \big),
\label{PQindependence}
\end{equation}

\noindent
or more generally

\begin{equation}
P \big( Q_{i,\odot} = q \big| i \in \G(S) \big) = P(Q_{i,\odot} = q),
\label{PQindependence}
\end{equation}

\noindent
substituting the specific distribution extracted from nodes in
$\G(S)$
by the general distribution over {\it all} nodes in the network.
The meaning is that while the statistical properties of 
$Q_i$
may, generally, depend on 
$S$,
with 
$Q(S) \ne Q(S^{\prime})$,
those of 
$Q_{i,\odot}$ 
are independent of 
$S$, 
providing
$Q_{\odot}(S) = Q_{\odot}(S^{\prime})$, 
ultimately providing
$Q_{\odot}(S) = \av Q_{\odot}$, an average over all nodes in the network.
This translates to 

\begin{equation}
Q_{\odot}(S) \equiv 
\frac{1}{|\G(S)|}\sum_{i \in \G(S)} \dfrac{1}{S} \sum_{n = 1}^N \m Ain Q_n = 
\dfrac{1}{N} \sum_{i = 1}^N \dfrac{1}{S_i} \sum_{n = 1}^N \m Ain Q_n \equiv 
\av Q_{\odot}
\label{CG}
\end{equation}

\noindent
where the l.h.s.\ represents a nearest neighbor average over nodes within 
$\G(S)$ 
and the r.h.s.\ represents a nearest neighbor average over all nodes, a characteristic of the 
{\it network}, independent of $S$.

\subsection{Steady state analysis}

We consider systems of the form (\ref{Dynamics}) that exhibit at least one fully positive steady state
$x_i$ ($i = 1,\dots,N$).
We focus on the dependence of this steady-state, 
$x_i$, 
on a node's weighted (incoming) degree
$S_i = \sum_{j = 1}^N A_{ij}$.
Therefore, we seek the average (time-dependent) activity 
$x(S,t)$ 
characterizing all nodes $i \in \G(S)$, which, substituting 
$x_i(t)$ for the random variable $Q_i$ in (\ref{Qs}), provides 

\begin{equation}
x(S,t) = \frac{1}{|\G(S)|}\sum_{i \in \G(S)} x_i(t).
\label{XRk}
\end{equation}

\noindent
Using (\ref{Dynamics}) we write

\begin{equation}
\dod {x(S,t)}{t} = 
\frac{1}{|\G(S)|}\sum_{i \in \G(S)}
\Big[
M_0 \big( x_i(t) \big) + \sum_{n = 1}^{N} \m Ain M_1 \big(x_i(t) \big) M_2 \big(x_n(t) \big)
\Big],
\end{equation}

\noindent
which we approximate by

\begin{equation}
\dod {x(S,t)}{t}  = 
M_0 \big( x(S,t) \big) + M_1 \big( x(S,t) \big)
\frac{1}{|G(S)|}
\sum_{i \in \G(S)}
\sum_{n = 1}^{N} \m Ain M_2 \big( x_n(t) \big).
\label{Ratek}
\end{equation}

\noindent
Equation (\ref{Ratek}) is exact in the limit where 

\begin{equation}
\frac{1}{|\G(S)|}\sum_{i \in \G(S)} M_{q} (x_i) 
\approx
M_{q} \left( \frac{1}{|\G(S)|}\sum_{i \in \G(S)} x_i \right),
\label{Mxkapproximation}
\end{equation}

\noindent
($q = 0,1$).
We can now use (\ref{CG}) to express the sum on the r.h.s.\ of (\ref{Ratek}) as

\begin{equation}
\frac{1}{|\G(S)|}
\sum_{i \in \G(S)}
\sum_{n = 1}^{N} \m Ain M_2 \big( x_n(t) \big) = 
S \av{M_2\big( x(t) \big)}_{\odot},
\label{NearestNeighborAv}
\end{equation}

\noindent
where $\av{M_2(x(t))}_{\odot}$, an average over all nearest neighbor nodes in the network, is independent of $S$.
Equation (\ref{Ratek}) then takes the form

\begin{equation}
\dod {x(S,t)}{t} = 
M_0 \big (x(S,t) \big ) + S M_1 \big ( x(S,t) \big ) \av{M_2(x(t))}_{\odot}.
\label{kiRate}
\end{equation}

\noindent
To obtain the steady state we set the l.h.s.\ of (\ref{kiRate}) to zero, providing

\begin{equation}
R \big ( x(S) \big ) = \frac{1}{\av{M_2(x)}_{\odot} S},
\label{Rx1overk}
\end{equation}

\noindent
where

\begin{equation}
R(x) = - \frac{M_1(x)}{M_0(x)}.
\label{R}
\end{equation}

\noindent
Extracting $x(S)$ from (\ref{Rx1overk}) we write

\begin{equation}
x(S) \sim R^{-1} \left( \lambda \right),
\label{Xk}
\end{equation}

\noindent
where 
$R^{-1}(x)$
is the inverse function of
$R(x)$
and

\begin{equation}
\lambda = \frac{1}{\av{M_2(x)}_{\odot} S} \sim S^{-1}
\label{lambda}
\end{equation}

\noindent
is the inverse weighted degree.
Equation (\ref{Xk}) expresses the average steady-state activity over all nodes with in-degree
$S$ ($i \in \G(S)$) in function of their inverted degree
$\lambda \sim S^{-1}$.


\subsection{The scaling of $\tau_i$}

We now calculate the response time 
$\tau_i$
of a node to a neighboring perturbation.
Hence, we induce a small permanent perturbation 
$\d x_m$ 
on the steady state activity of node 
$m$,
a nearest neighbor of 
$i$, 
setting

\begin{equation}
x_m(t) = x_m + \d x_m.
\label{Xnt}
\end{equation}

\noindent
The dynamic equation (\ref{Dynamics}) then becomes 

\begin{eqnarray}
\frac{\d }{\d t}(x_i + \d x_i) &=& 
M_0(x_i + \d x_i) + 
\sum_{\substack{j = 1 \\ j \ne m}}^{N} 
\m Aij M_1(x_i + \d x_i)M_2(x_j + \d x_j) 
\nonumber \\
&+& \m Aim M_1(x_i + \d x_i)M_2(x_m + \d x_m),
\label{PerturbedRate}
\end{eqnarray}

\noindent
where $\d x_i$ and $\d x_j$ ($j = 1,\dots,N, j \ne m$) are all time dependent, while
$\d x_m$ is constant.
Linearizing around the steady state we obtain

\begin{eqnarray}
\frac{\d}{\d t} (\d x_i) &=& 
\left( 
M_0^{\prime}(x_i) + M_1^{\prime}(x_i) \sum_{j = 1}^{N} \m Aij M_2(x_j) 
\right) \d x_i(t) 
\nonumber \\
&+& 
M_1(x_i) \sum_{j = 1}^{N} \m Aij M_2^{\prime}(x_j) \d x_j(t) + 
O(\d x^2),
\label{LinearRate}
\end{eqnarray}

\noindent
where
$M_q^{\prime}(x)$ ($q = 0,1,2$)
represents the derivative
$\d M_q/\d x$
with
$x$
taken at the steady state, which according to (\ref{Xk}) can be expressed by
$x = R^{-1}(\lambda)$.
Next, following a similar derivation as the one leading to (\ref{kiRate}), 
we average of over all nodes in 
$\G(S)$
to obtain a direct equation for the response of nodes with weighted degree 
$S$

\begin{equation}
\d x(S,t) = \dfrac{1}{|\G(S)|} \sum_{i \in \G(S)} \dif x_i(t).
\label{dxS}
\end{equation} 

\noindent
Using (\ref{LinearRate}) to express the time derivative of $\d x_i(t)$ in (\ref{dxS})
and neglecting the higher order terms $O(\d x^2)$, we obtain

\begin{eqnarray}
\frac{\d}{\d t}\big( \dif x(S,t) \big) &=& 
\left( 
M_0^{\prime}\big( x(S) \big) + M_1^{\prime}\big( x(S) \big) 
\dfrac{1}{|\G(S)|}\sum_{i \in \G(S)} \sum_{j = 1}^{N} \m Aij M_2(x_j) 
\right) \d x(S,t) 
\nonumber \\ &+& 
M_1 \big( x(S) \big)  
\dfrac{1}{|\G(S)|}\sum_{i \in \G(S)} \sum_{j = 1}^N \m Aij M_2^{\prime}(x_j) \d x_j(t),
\label{ddxSdt}
\end{eqnarray}
 
\noindent
where 
$x(S)$ is the steady state activity of nodes in $\G(S)$, as expressed in (\ref{Xk}).
Finally, the configuration model assumption, allows us to simplify the first sum on the r.h.s.\ 
using (\ref{CG}), providing us with

\begin{eqnarray}
\frac{\d}{\d t}\big( \dif x(S,t) \big) &=& 
\left( 
M_0^{\prime}\big( x(S) \big) + S M_1^{\prime}\big( x(S) \big) 
\av {M_2(x)}_{\odot} \right) \d x(S,t) + f(S,t), 
\label{ddxSdtCG}
\end{eqnarray}

\noindent
where

\begin{equation}
f(S,t) = M_1 \big( x(S) \big)  
\dfrac{1}{|\G(S)|}\sum_{i \in \G(S)} \sum_{j = 1}^N \m Aij M_2^{\prime}(x_j) \d x_j(t).
\label{fSt}
\end{equation}

\noindent
Equation (\ref{ddxSdtCG}) can be written in the form

\begin{equation}
\frac{\d}{\d t} \big( \d x(S,t) \big) = -\frac{1}{\tau(S)} \d x + f(S,t),
\label{LinearRateCollected}
\end{equation}

\noindent
in which the average relaxation time $\tau(S)$ follows

\begin{equation}
\frac{1}{\tau(S)} = 
M_0^{\prime} \big( x(S) \big) + S M_1^{\prime} \big( x(S) \big) \av {M_2(x)}_{\odot}.
\label{1overTaui}
\end{equation}

\noindent
Equation (\ref{LinearRateCollected}) is a non-homogeneous linear differential equation,
describing the average time dependent response
$\d x(S,t)$
of nodes in $\G(S)$ 
to a neighboring permanent perturbation
$\d x_m$.
Its solution takes the form

\begin{equation}
\d x(S,t) = 
C e^{-\frac{t}{\tau(S)}} + 
e^{-\frac{t}{\tau(S)}}
\int_{0}^{t} f(S,t^{\prime})
e^{\frac{t^{\prime}}{\tau(S)}}
\d t^{\prime},
\label{dxit}
\end{equation}

\noindent
where the constant
$C$
is set to zero to satisfy the initial condition
$\d x(S,t = 0) = 0$.
The relaxation of 
$\d x(S,t)$ (\ref{dxit})
to its final, perturbed, state is governed by
$\tau(S)$ (\ref{1overTaui}),
which depends on the weighted degree 
$S$, both explicitly, and implicitly through $x(S)$ in (\ref{Xk}).
To observe this we focus on each of the two terms on the r.h.s.\ of (\ref{1overTaui}) independently.
First we write

\begin{equation}
M_0^{\prime} \big( x(S) \big) = 
\left.
\frac{\d M_0}{\d x}
\right|_{x = R^{-1}(\lambda)},
\label{M0prime}
\end{equation}

\noindent
a derivative around the steady state
$x(S)$,
which we expressed using (\ref{Xk}).
Using the definition of 
$R(x)$ 
(\ref{R}) we further develop (\ref{M0prime}) and write

\begin{eqnarray}
M_0^{\prime} \big( x(S) \big) &=& 
\left.
\left(-
\dfrac{M_1^{\prime}(x)}{R(x)} +
\dfrac{M_1(x)}{R^2(x)} R^{\prime}(x)
\right)
\right|_{x = R^{-1}(\lambda)} 
\nonumber \\
\nonumber \\
&=& 
-\dfrac{M_1^{\prime} \big( R^{-1}(\lambda) \big)}
{\lambda} + 
\dfrac{M_1 \big( R^{-1}(\lambda) \big) R^{\prime} \big( R^{-1}(\lambda) \big)}
{\lambda^{2}},
\end{eqnarray}
 
\noindent
where in the last step we used 
$R(R^{-1}(\lambda)) = \lambda$.
In a similar fashion we express the second term of 
(\ref{1overTaui}) as

\begin{equation}
S M_1^{\prime} \big ( x(S) \big) \av{M_2(x)}_{\odot} =
\av{M_2(x)}_{\odot} 
\dfrac{M_1^{\prime} \big( R^{-1}(\lambda) \big)}{\lambda}.
\end{equation}

\noindent
Collecting all the terms we arrive at

\begin{equation}
\frac{1}{\tau(S)} \sim
c_1
\dfrac{M_1^{\prime} \big( R^{-1}(\lambda) \big)}{\lambda} +
c_2 
\dfrac{M_1 \big( R^{-1}(\lambda) \big) R^{\prime} \big( R^{-1}(\lambda) \big)}
{\lambda^{2}},
\label{1overTauS}
\end{equation}

\noindent
where the coefficients are

\begin{eqnarray}
c_1 &=& 1 - \av{M_2(x)}_{\odot}
\nonumber \\
c_2 &=& -1.
\end{eqnarray}

\noindent
As we are only interested in the scaling of 
$\tau(S)$
with
$S$ (or $\lambda$)
in the limit of large 
$S$ (small $\lambda$), 
we can rewrite (\ref{1overTauS}) without the coefficients.
Indeed, for sufficiently large $S$, only the leading terms where
$S$
is raised to the highest power dominate the equation,
providing
$1/\tau(S) \sim c_1 S^a + c_2 S^b \sim S^{\max(a,b)}$,
independent of 
$c_1$ and $c_2$. 
Hence, preserving only the terms relevant to the scaling, Eq. (\ref{1overTauS}) becomes

\begin{eqnarray}
\frac{1}{\tau(S)} &\sim&
\frac{1}{\lambda^2}
\Big[ 
R \left( R^{-1}(\lambda) \right) M_1^{\prime} \left( R^{-1}(\lambda) \right) 
+
M_1 \left( R^{-1}(\lambda) \right) R^{\prime} \left( R^{-1}(\lambda) \right)
\Big]
\nonumber \\[5pt]
&=& 
\frac{1}{\lambda^2}
\left.
\frac{\d}{\d x}\big( M_1(x)R(x) \big)
\right|_{x = R^{-1}(\lambda)},
\end{eqnarray}

\noindent
where, once again, we used 
$\lambda = R(R^{-1}(\lambda))$,
leading to the extracted pre-factor of 
$\lambda^{-2}$.
We can now write

\begin{equation}
\tau(S) \sim
\lambda^2
Y\left( R^{-1}(\lambda) \right)
\label{taulambda}
\end{equation}

\noindent
where

\begin{equation}
Y(x) = 
\left(
\frac{\d (M_1R)}{\d x}
\right)^{-1}.
\label{Y}
\end{equation}

\noindent
Equation (\ref{taulambda}) expresses 
$\tau(S)$
as a function of 
$\lambda$ (\ref{lambda}),
from which its dependence on 
$S$
can be obtained.
It indicates that the scaling of 
$\tau(S)$
with 
$S$
is determined directly by the dynamical functions
$M_1(x)$
and
$R(x)$,
or, using (\ref{R}),
$M_1(x)$
and
$M_0(x)$.
Next we express  
$Y(R^{-1}(\lambda))$
as a Hahn series \cite{Hahn1995} around $\lambda = 0$

\begin{equation}
Y\big ( R^{-1}(\lambda) \big ) = 
\sum_{n = 0}^{\infty} C_n \lambda^{\Gamma(n)}, 
\label{YLaurent}
\end{equation}

\renewcommand{\O}{\mathcal O}

\noindent
allowing us to systematically consider the asymptotic behavior at 
$S \rightarrow \infty$, equivalent to $\lambda \rightarrow 0$.
The Hahn series is a generalization of the Taylor expansion to allow for both negative and real powers,
as represented by  
$\Gamma(n)$,
a countable set of real numbers, ordered such that
$\Gamma(n - 1) < \Gamma(n) < \Gamma(n + 1)$.
Hence the leading power of (\ref{YLaurent}) is 
$\Gamma(0)$, the next leading power is $\Gamma_R(1)$, etc..
For large $S$ we only keep the leading order term,
namely 
$\lambda^{\Gamma(0)}$.
This provides us with (\ref{taulambda})

\begin{equation}
\tau(\lambda) \sim \lambda^2 \lambda^{\Gamma(0)},
\label{tlambda}
\end{equation}

\noindent
or, substituting $S^{-1}$ for $\lambda$, 

\begin{Frame}
\begin{equation}
\tau(S) \sim S^{\theta},
\label{tauk}
\end{equation}

\noindent
where 

\begin{equation}
\theta = -2 - \Gamma(0),
\label{theta}
\end{equation}
\end{Frame}

\noindent
as presented in Eqs.\ (2) - (4) in the main paper text.


\section{Classification of the dynamic models}
\label{Models}

We analyzed the propagation patterns in six different 
frequently used dynamic models, for each obtaining 
$\theta$ (\ref{theta}), and hence their class as
distance driven ($\theta = 0$),
degree driven ($\theta > 0$)
or composite ($\theta < 0$).
The detailed derivations appear below.

\subsection{Regulatory dynamics - $\R_1$ and $\R_2$}

Gene regulation is often modeled using Michaelis-Menten dynamics,
in which the activity, {\it i.e.} expression, of all genes follows \cite{Alon2006,Karlebach2008}, 
   
\begin{eqnarray}
\dod {x_i}{t} &=& -Bx_i^a+\sum_{j = 1}^N  \m Aij  \mathcal{H}(x_j) 
\label{Chapter1_Regulatory _Dynamics_eqn1},
\end{eqnarray}

\noindent
where 
$\mathcal{H} (x_j)$ 
is the Hill function describing the  activation/inhibition of 
$x_i$ by $x_j$.  
Since regulation depends primarily on the presence or absence of
$x_j$, 
with little sensitivity to $j$'s specific abundance, the Hill function is designed to be a \emph{switch-like} function satisfying
$\mathcal{H}(x_{j}) \rightarrow 1$ ($\mathcal{H}(x_j)\rightarrow 0$) 
for large (small) $x_j$ in case 
$x_j$ activates $x_i$, or 
$\mathcal{H}(x_j)\rightarrow 1$ ($\mathcal{H}(x_j)\rightarrow 0$) 
for small (large) 
$x_j$ in the case of inhibition. A most common choice is \cite{Alon2006,Karlebach2008} 

\begin{eqnarray}
\dod {x_i}{t} &=& -Bx_i^a + \sum_{j = 1}^N \m Aij \frac{x_j^h}{1 + x_j^h},
\label{MM}
\end{eqnarray}

\noindent
where the Hill coefficient $h$ governs the rate of saturation of
$\mathcal{H} (x_j)$. 
Equation (\ref{MM}) can be cast in the form (\ref{Dynamics}) with
$M_0(x) = -Bx^a$, $M_1(x) = 1$ and $M_2(x) = x^h/(1 + x^h)$. 
Hence 
$R(x)$ (\ref{R})
becomes
$R(x) = - B^{-1} x^{-a}$,
and its inverse follows

\begin{equation}
R^{-1}(x) = B^{-\frac{1}{a}} x^{-\frac{1}{a}} \sim x^{-\frac{1}{a}}.
\label{Trans_Regulatory_Dynamics_eqn3}
\end{equation}

\noindent
Next we use (\ref{Y}) to write

\begin{equation}
Y(x)= \left( \dod {M_1R}{x} \right)^{-1},
\end{equation}

\noindent
which taking the above $R(x)$ becomes

\begin{equation}
Y(x) = \left(\dod {x^{-a}}{x} \right)^{-1} \sim x^{(a+1)}.
\label{Trans_Regulatory_Dynamics_eqn5}  
\end{equation}

\noindent
Using (\ref{Trans_Regulatory_Dynamics_eqn3}) in (\ref{Trans_Regulatory_Dynamics_eqn5}) we arrive at the Hahn expansion of (\ref{YLaurent}) 

\begin{equation}
Y \big( R^{-1} (\lambda) \big )
\sim Y \big(\lambda^{-\frac{1}{a}} \big) \sim \lambda^{-\frac{a + 1}{a}},
\label{Trans_Regulatory_Dynamics_eqn6}
\end{equation}

\noindent
whose leading (indeed, only) power is
$\Gamma(0) = -(a + 1)/a$.
Finally, we predict 
$\theta$ from (\ref{theta}) to be

\begin{Frame}
\begin{equation}
\theta = -2 - \Gamma(0) = -2 + \frac{a+1}{a} = \frac{1 - a}{a}.
\label{Trans_Regulatory_Dynamics_eqn7}
\end{equation}
\end{Frame}
 
\noindent
For $\R_1$ we set $a = 1$ and $h = 1$, predicting 
$\theta = 0$, a distance driven dynamics;
for $\R_2$ we set $a = 0.4$ and $h = 0.2$, predicting 
$\theta = 3/2$, a degree driven system. 
Both predictions are perfectly confirmed on both model and real networks in 
Fig.\ 3 of the main text.

\subsection{Population dynamics - $\P_1$ and $\P_2$}

Birth-death processes have many applications in population dynamics \cite{Novozhilov2006}, queuing theory \cite{Hayes2004} or biology \cite{Novozhilov2006}. We consider a network in which the nodes represent sites, each site $i$ having a population $x_i$, where population flow is enabled between neighboring sites. 
This process can be described by 

\begin{equation}
\dod {x_i}{t} = -Bx_i^b + \sum_{j = 1}^{N} \m Aij x_j^a,
\label{Trans_PD_Dynamics_eqn1}
\end{equation}
 
\noindent 
where the first term on the r.h.s.\ represents the internal dynamics of site 
$i$, 
characterized by the exponent 
$b$, 
which distinguished between processes such as \cite{Barzel2011} 
in/out flux ($b = 0$),
mortality ($b = 1$),
pairwise annihilation ($b = 2$), etc..
The second term describes the nonlinear flow from 
$i$'s neighboring sites 
$j$ into
$i$.
Here we have
$M_0(x) = -B x^b$, $M_1(x) = 1$ and $M_2(x) = x^a$, 
therefore  
$R(x) = - B^{-1} x^{-b}$.
Following the same steps leading from 
(\ref{Trans_Regulatory_Dynamics_eqn3}) to (\ref{Trans_Regulatory_Dynamics_eqn6}) we find   
 
\begin{eqnarray}
Y \big( R^{-1}(\lambda) \big) \sim \lambda^{-\frac{b + 1}{b}},
\label{Trans_PD_Dynamics_eqn4}
\end{eqnarray}
  
\noindent
predicting
 
\begin{Frame}
\begin{eqnarray}
\theta = -2 - \Gamma(0) \sim -2 + \frac{b + 1}{b}.
\label{Trans_PD_Dynamics_eqn5}
\end{eqnarray}
\end{Frame}
 
\noindent
For $\P_1$ we set $b = 1$ and $a = 0.25$, predicting 
$\theta = 0$, a distance driven dynamics;
for $\P_2$ we set $b = 0.5$ and $a = 0.2$, predicting 
$\theta = 1$, a degree driven system, 
both in perfect agreement with Fig.\ 3 of the main text.

\subsection{Epidemics - $\E$}
   
In the susceptible-infected-susceptible (SIS) model, each node may be in one of two potential states: infected 
($I$) or susceptible ($S$). 
The spreading dynamics is driven by the two process 
   
\begin{equation}
I + S \rightarrow 2I,
\label{chapter1_SIS_Dynamics_eqn1}
\end{equation}
   
\noindent   
where a susceptible node becomes infected by contact with one of its infected neighbors, and

\begin{equation}
I \rightarrow S,
\label{chapter1_SIS_Dynamics_eqn2}
\end{equation}
   
\noindent
an infected node recovering and becoming susceptible again. 
The activity 
$x_i(t)$
denotes the probability that $i$ is in the infected state.
The infection and recovery processes above can be captured by
\cite{Dodds2005}

\begin{equation}
\dod {x_i}{t} = -Bx_i + \sum_{j = 1}^N \m Aij (1 - x_i)x_j.\label{Trans_SIS_Dynamics_eqn1}
\end{equation}
   
\noindent
The first term on the r.h.s.\ accounts for the process of recovery and the second term accounts for the process of infection, where a node could only  become infected if its in the susceptible state, with probability 
$1 - x_i$,
and its neighbor is in the infected state, with probability 
$x_j$.
We have 
$M_0(x) = -Bx$, $M_1(x) = 1 - x$ and $M_2(x) = x$,
providing (\ref{R})
   
\begin{equation}
R(x) = \frac{1-x}{Bx},     
\label{Trans_SIS _Dynamics_eqn2}
\end{equation}

\noindent
and therefore

\begin{equation}
R^{-1}(x) = \frac{1}{1 + Bx}.
\label{RinvSIS}
\end{equation}
   
\noindent
Equation (\ref{Y}) takes the form   

\begin{equation}
Y(x) = \left( \dod{}{x}
\left( \frac{(1 - x)^2}{x} \right) 
\right)^{-1} \sim \frac{1}{1 - x^{-2}},
\label{Trans_SIS_Dynamics_eqn3}  
\end{equation}
   
\noindent
allowing us to obtain the Hahn expansion (\ref{YLaurent}) as 
    
\begin{equation}
Y \big( R^{-1}(\lambda) \big) =
Y \left( \frac{1}{1 + B \lambda} \right) \sim 
\frac{1}{1 - (1 + B\lambda)^2} \sim 
\dfrac{1}{2B} \lambda^{-1} + \dfrac{1}{4}\lambda^0 + \dfrac{1}{8} B\lambda^1 + O(\lambda^2),
\label{Trans_SIS_Dynamics_eqn4}
\end{equation}
   
\noindent
whose leading power is 
$\Gamma(0) = -1$.
Using (\ref{theta}) this predicts

\begin{Frame}
\begin{equation}
\theta = -2 - \Gamma(0) = -2 - (-1) = -1,
\label{Trans_SIS_Dynamics_eqn5}
\end{equation}
\end{Frame}   
   
\noindent
a composite dynamics, in which hubs respond most rapidly (Fig.\ 3 in main paper).

\subsection{Mutualistic dynamics in ecology - $\M$}  

We consider symbiotic eco-systems, such as plant-pollinator networks, in which the interacting species exhibit symbiotic relationships. 
The species populations follow the dynamic equation 

\begin{equation}
\dod {x_i}{t} = B x_i(t) \left(1 - \frac{x_i^a(t)}{C} \right)
+ \sum\limits_{j = 1}^N A_{ij} x_i(t)F\big(x_j(t)\big).
\label{Ecology_dynamics_eq}
\end{equation}  

\noindent
The self dynamics

\begin{equation}
M_0(x) = Bx \left(1 - \frac{x^a}{C} \right)
\end{equation}

\noindent
is a generalization of the frequently used logistic growth: 
when the population is small, the species 
reproduce at a rate 
$B$, yet, as $x_i$ approaches the carrying capacity of the system $C$, growth is hindered by competition over limited resources \cite{May1976b},
captured by the nonlinear 
$-x_i^{a + 1}$ term.
For 
$a = 1$ we arrive at the classic quadratic growth deficiency term, 
in which competition scales with the number of competing 
{\it pairs}.
In case $a > 1$ growth is hindered through higher order 
internal competition within a species. 

The mutualistic inter-species interactions are captured by

\begin{equation}
    \begin{array}{ll}
        M_1(x) = x\\
        M_2(x) = F(x),
    \end{array}
\end{equation}

\noindent
where $F(x)$ represents the {\it functional response}, describing the positive impact that 
species $j$ has on species $i$.
This functional response can take one of several forms \cite{Holling1959}:

\noindent \textbf{Type \Rom{1}:} linear impact 
    \begin{equation}
        F(x) = \alpha x.
        \label{def:Type1}
    \end{equation}
\noindent \textbf{Type \Rom{2}:} saturating impact 
    \begin{equation}
        F(x) = \frac{\alpha x}{1 + \alpha x}.
        \label{def:Type2}
    \end{equation}
\noindent \textbf{Type \Rom{3}:} A generalization of Type \Rom{2}, where 
    \begin{equation}
        F(x) = \frac{\alpha x^h}{1 + \alpha x^h}.
        \label{def:Type3}
    \end{equation}

\noindent
In our simulations we used Type \Rom{2} mutualistic interactions
and set the competition term to $a = 2$, providing

\begin{equation}
\begin{array}{rcl}
M_0(x) &=& Bx\left(1 - \dfrac{x^2}{C}\right) 
\\
M_1(x) &=& x
\\[8pt]
M_2(x) &=& \dfrac{\alpha x}{1 + \alpha x},
\label{mutualistic_functions}
\end{array}
\end{equation}

\noindent
where, for simplicity, we set $B = C = \alpha = 1$.
Hence we have (\ref{R})

\begin{equation}
R(x) = \dfrac{1}{1 - x^2},
\end{equation}

\noindent
and therefore

\begin{equation}
R^{-1}(x) \sim \left( \frac{x - 1}{x} \right)^{\frac{1}{2}}.
\end{equation}

\noindent
Next we use (\ref{Y}) to write
    
\begin{equation}
Y(x) = \left( \dod {}{x} 
\left (
\frac{x}{1 - x^2} 
\right) \right)^{-1} = \dfrac{\left( 1 - x^2 \right)^2}{1 + x^2}.
\label{Trans_ECO_Dynamics_eqn3}  
\end{equation}
 
\noindent    
Consequently, the Hahn expansion (\ref{YLaurent}) takes the form

\begin{eqnarray}
Y \big( R^{-1}(\lambda) \big) = 
Y \left( \left( \dfrac{\lambda - 1}{\lambda} \right)^{\frac{1}{2}} \right) =
\dfrac{1}{2\lambda^2 - \lambda} = 
-\lambda^{-1} - 2\lambda^{0} + 5\lambda^{1} + O(\lambda^{2}),
\label{Trans_ECO_Dynamics_eqn4}
\end{eqnarray}

\noindent
for which the leading power
$\Gamma(0) = -1$.
As a result we predict (\ref{theta})

\begin{Frame}      
\begin{eqnarray}
\theta = -2 - \Gamma(0) = -1,
\label{Trans_ECO_Dynamics_eqn5}
\end{eqnarray}
\end{Frame}
       
\noindent
classifying $\M$ in the composite dynamics class, as
fully confirmed by the results presented in Fig.\ 3 of the main paper.

\begin{table}
\centering
\includegraphics[width=1\textwidth]{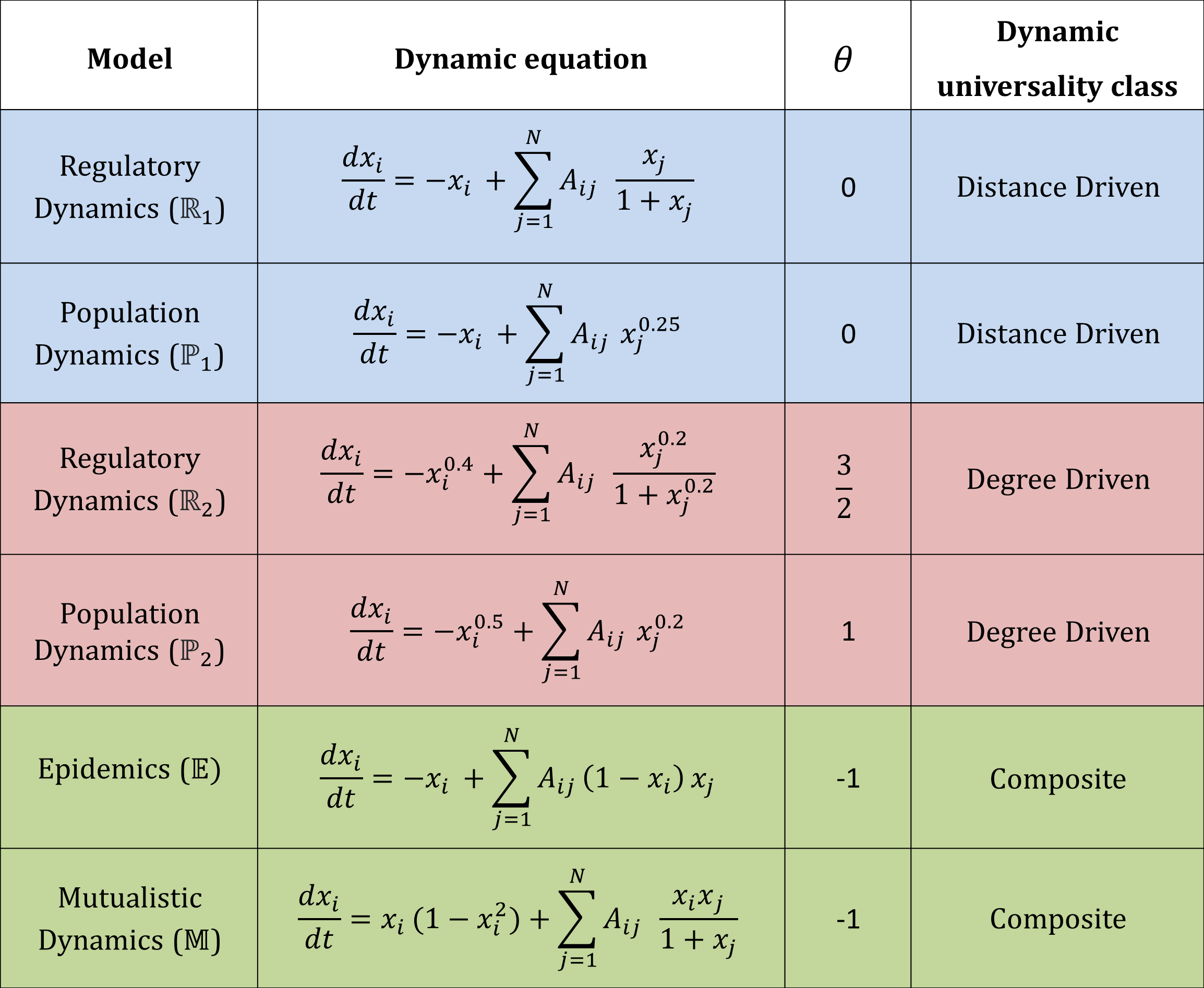}
\caption{
\textbf{Dynamic models.} Summary and classification of all dynamic models.}
\label{fig:table}
\end{table}
       
\pagebreak

\section{Methods and data analysis}
\label{Supplementary Information 3: Methods and Data Analysis}
\renewcommand{\d}{{\rm d}}

\subsection{Numerical integration}
\label{numerical_inte}

To numerically test our predictions we constructed Eq.\ (\ref{Dynamics})
for each of the systems in Table \ref{fig:table}, using
the appropriate 
$\m Aij$ (Scale-free, Erd\H{o}s-R\'{e}nyi, empirical, etc.).
We then used a fourth-order Runge-Kutta stepper (Matlab's ode45) 
to numerically solve the resulting equations. 
Starting from an arbitrary initial condition
$x_i(t = 0)$, $i = 1,\dots,N$ 
we allowed the system to reach its steady state by waiting for 
$\dot x_i \rightarrow 0$.
To numerically realize this limit we implemented the termination condition 

\begin{equation}
\max_{i = 1}^N \left| \frac{x_i(t_n) - x_i(t_{n - 1})}{x_i(t_n) \Delta t_n}\right| < \varepsilon,
\label{TerminationCondition}
\end{equation}
    
\noindent
where $t_n$ is the time stamp of the $n$th Runge-Kutta step and
$\Delta t_n = t_n - t_{n - 1}$.
As the system approaches the steady-state, the activities
$x_i(t_n)$
become almost independent of time, and the numerical derivative
$\dot x_i = x_i(t_n) - x_i(t_{n - 1}) / \Delta t_n$
becomes small compared to
$x_i(t_n)$.
The condition (\ref{TerminationCondition}) guarantees that the maximum of 
$\dot x_i/x_i$ 
over all activities
$x_i(t_n)$
is smaller than the pre-defined termination variable
$\varepsilon$.
Across the six systems we tested we set $\varepsilon \le 10^{-12}$, 
a rather strict condition, to ensure that our system is sufficiently close to the {\it true} steady state.

\subsection{Measuring $\T ij$ and $\tau_i$}
    
To observed the spatio-temporal propagation of a perturbation we set the initial condition of the 
system to its numerically obtained steady-state above. 
We then introduce a boundary condition on the source node $j$, as 

\begin{eqnarray}
x_j(t) = x_j + \Delta x_j,  
\label{chapter2perb1}   
\end{eqnarray} 
      
\noindent
a signal in the form of a permanent perturbation to $j$'s steady state activity
$x_j$.
In our simulated results we used
$\Delta x_j = \alpha x_j$, setting $\alpha = 0.1$, a $10 \%$ perturbation.
The remaining 
$N - 1$
nodes continue to follow the original dynamics (\ref{Dynamics}), responding to the
propagating signal 
$\Delta x_j$.
To be explicit, we simulate this propagating perturbation by numerically solving the {\it perturbed} Eq.\ (\ref{Dynamics}), which now takes the form

\begin{equation}
\left\{
  \begin{array}{rcll}
    \dod{x_j}{t} &=& 0 & 
    \\ \\
    \dod{x_i}{t} &=& M_0(x_i) + \sum_{n = 1}^N A_{ij} M_1(x_i)M_2(x_n) & i \ne j
  \end{array}
\right.,
\label{PerturbedDynamics}
\end{equation}

\noindent
in which the perturbation on $j$ is held constant in time, and the remaining $N - 1$ nodes
respond via the system's intrinsic dynamics.
The system's response is then obtained as

\begin{equation}
x_i(t) = x_i + \Delta \m x ij(t),
\label{Response}
\end{equation}

\noindent
in which
$\Delta \m x ij(t)$
represents 
$i$'s temporal response to the signal $\Delta x_j$.
We continue running (\ref{PerturbedDynamics}) until the termination condition 
(\ref{TerminationCondition}) is realized again,
and the system reaches its new perturbed state with
$\Delta \m x ij(t \rightarrow \infty) = \Delta \m x ij$,
$i$'s final response to $j$'s signal.
To focus on the response time of each node, we define $i$'s normalized response as

\begin{equation}
\m f ij(t) = \dfrac{\Delta \m x ij(t)}{\Delta \m x ij},
\label{fijt}
\end{equation}

\noindent
which transitions smoothly between
$\m f ij(t) = 0$ at $t = 0$ to $\m f ij(t) = 1$ at $t \rightarrow \infty$,
as $i$ approaches its final response.
The function $\m f ij(t)$ captures the spatio-temporal response of the system in the discrete 
network space, namely the level of response obtained at time $t$ in location $i$.
When $\m f ij(t) = \eta$, we say that $i$ has reached an $\eta$-fraction of its final response
to the traveling signal $\dif x_j$.
For instance, setting $\eta = 1/2$ allows us to evaluate the {\it half-life} of $i$'s response. 
This allows us to evaluate the propagation time $\T ij$ as the time when

\begin{equation}
\m f ij \big( t = \T ij \big) = \eta
\label{fiTij}
\end{equation}

\noindent
or alternatively

\begin{equation}
\T ij = \m f ij^{-1} (\eta).
\label{Tij}
\end{equation}

\begin{figure}[h!]
	\centering
	\includegraphics[width=0.65\textwidth]{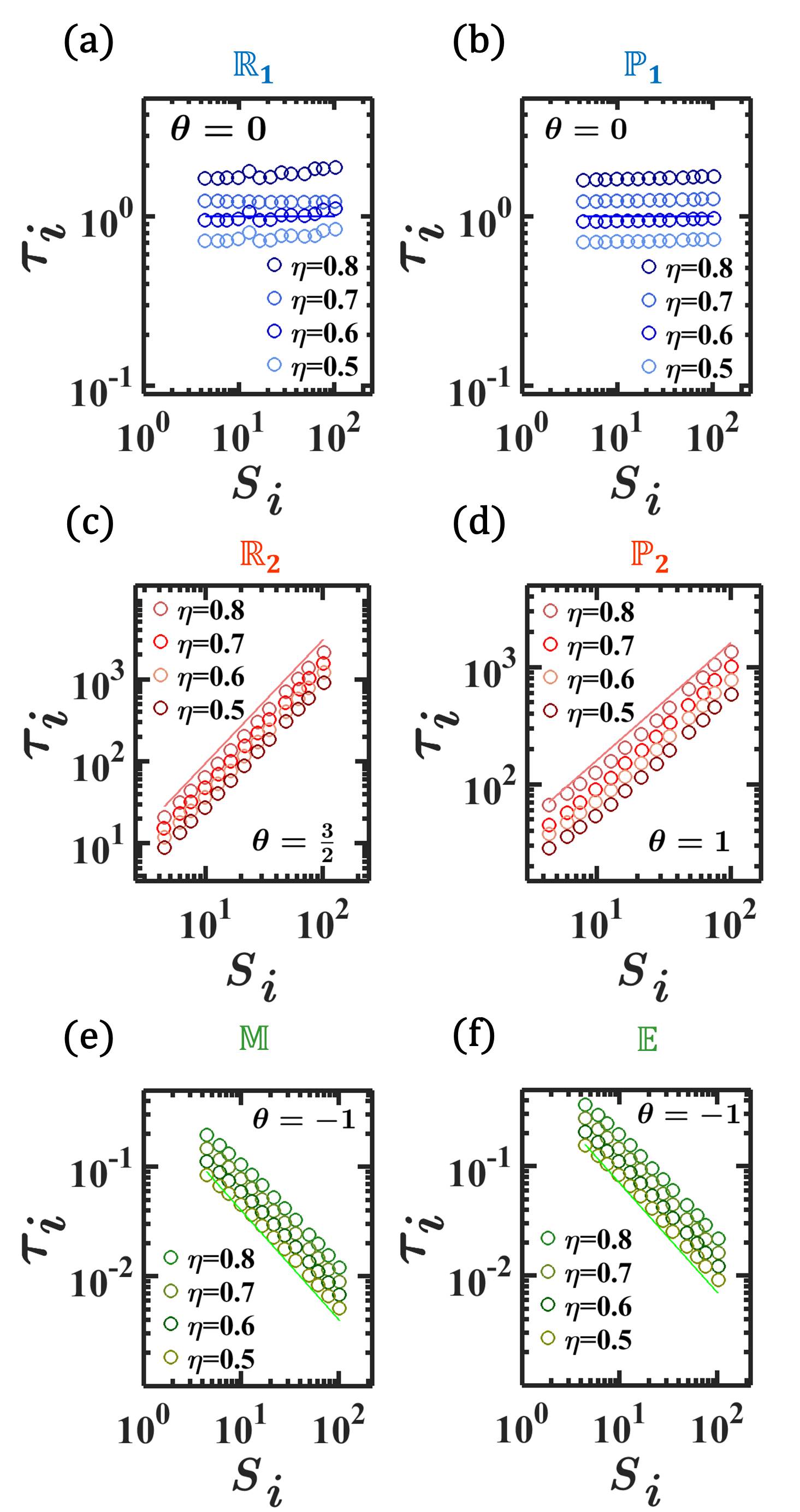}
	\caption{
		\textbf{\bf The impact of the arbitrary parameter $\eta$}.
		Measuring the response times requires to select the value of 
		$\eta$ in (\ref{Tij}), an arbitrary parameter between zero and 
		unity. To test the impact of this parameter we tested our results 
		for the scaling $\theta$ (\ref{tauk}) under different selected 
		values of $\eta$. As expected, we find that our results are not 
		affected by $\eta$.  	 
}
	\label{FigEta}
\end{figure}

\noindent
The parameter $\eta$ can be set to any value between zero and unity, 
$\eta \in (0,1)$,
with the typical choice being of order 
$\eta \sim 1/2$.
All results presented in the main paper were obtained for $\eta = 0.7$,
however, as we show in Fig.\ \ref{FigEta} changing the value of 
$\eta$ has no detectable effect on the observed behavior of $\tau_i$
and hence of $\T ij$.

\noindent
{\bf Local response}. To obtain the local response
$\tau_i$
we must measure the response time to a signal in the direct vicinity of $i$,
namely $\T ij$ where $j$ is directly linked to $i$.
Hence we denote by $K_i$ the group of incoming neighbors of $i$

\begin{equation}
K_i = \{j = 1,\dots,N | \m Aij \ne 0 \}
\label{Ki}
\end{equation}

\noindent 
and average $i$'s response time to these neighbors as

\begin{equation}
\tau_i = \dfrac{1}{|K_i|} \sum_{j \in K_i} \T ij,
\label{Def:taui}
\end{equation}    

\noindent
where $|K_i|$ is the number of nodes in $K_i$.
    
\noindent
{\bf Visualizing the spatio-temporal spread.}
To construct the visualizations of Fig.\ 2a - f in the main paper we used 
Gephi \cite{Bastian2009} to layout the weighted scale-free network SF, placing the source node 
$j$ at the center.
As the propagation unfolds we measured the response of all nodes 
$i = 1,\dots,N$,
setting the size and color depth of each node to be linearly proportional to 
$\m f ij(t)$ (\ref{fijt}).
Later, in Fig.\ 5 of the main text we present the exact same data only this time we laid out the nodes according to our universal metric $\L ij$,
as described in Eq.\ (5) of the main text.
Hence we located $j$ at the center as before ($\L jj = 0$), and placed all target nodes $i$ at a radial distance $r$
proportional to $\L ij$, with randomly 
selected azimuth $\varphi \in [0,2\pi]$.  
In the case of degree-driven propagation ($\theta > 0$, red)
since $\L ij$ (and $\T ij$) span several orders of magnitude we set
$r \sim \ln \L ij$.

\subsection{Logarithmic binning}
   
The scaling $\tau_i \sim S_i^{\theta}$ is shown in log-scale in Fig.\ 3 
of the main paper, with $\theta$ captured by the linear slope of 
$\tau_i$ vs.\ $S_i$. 
To construct these plots we employed logarithmic binning \cite{Milojevic2010}.
First we divide all nodes into $W$ bins
      
\begin{equation}
\B(w) = \big\{ i = 1,\dots,N \big{|} c^{w-1} < S_i \le c^w \big\},
\label{wbinning}
\end{equation}

\noindent
where
$w=1,...,W$ and 
$c$ is a constant. In (\ref{wbinning}) the $w$th bin includes all nodes $i$ whose weighted degrees $S_i$ are between 
$c^{w-1}$ and $c^{w}$. The parameter $c$ is selected
such that the unity of all bins 
$\cup_{w = 1}^W \B(w)$ includes all nodes, hence we set
$c^{W} = \max{S_i}$.
We then plot the average degree of the nodes in each bin 

\begin{equation} 
S_w = \langle S_i \rangle_{i \in \B(w)} = 
\dfrac{1}{|\B(w)|}\sum_{i \in \B(w)} S_i
\label{SwBinning}
\end{equation}

\noindent
versus the average response time of nodes in that bin

\begin{equation}    
\tau_w = \langle \tau_i\rangle _{i \in \B(w)} = 
\dfrac{1}{|\B(w)|}\sum_{i \in \B(w)} \tau_i.
\label{twBinning}
\end{equation}

\noindent
To evaluate the measurement error for each bin we first calculated the variance in the observed $\tau_i$ across all nodes in the bin 
$\sigma_w^2 = \langle \tau_i^2 \rangle_{i \in \B(w)} - \langle \tau_i \rangle ^2_{i \in \B(w)}$. We then set the error-bar to represent the $95\%$ confidence interval as \cite{Cox1974}
    
\begin{equation}
E_w = \frac{1.96 \sigma_w}{\sqrt{|\B(w)|}}.
\label{chapter4binning2}
\end{equation}

\noindent
A similar scheme was used to present
$\av {\T ij}$ vs.\ $N$ in Fig.\ 4d - f and 
$\T ij$ vs.\ $\L ij$ in Fig.\ 5h - j of the main paper.
In most cases the error bars were tiny, smaller than the size of the plot markers. 

\subsection{Model and empirical networks}
\label{Networks}

To test our predictions we constructed several model and real networks 
with highly diverse topological characteristics, as summarized below:

   \begin{table}[h]
   	\centering
   	
   	\includegraphics[width=1\textwidth]{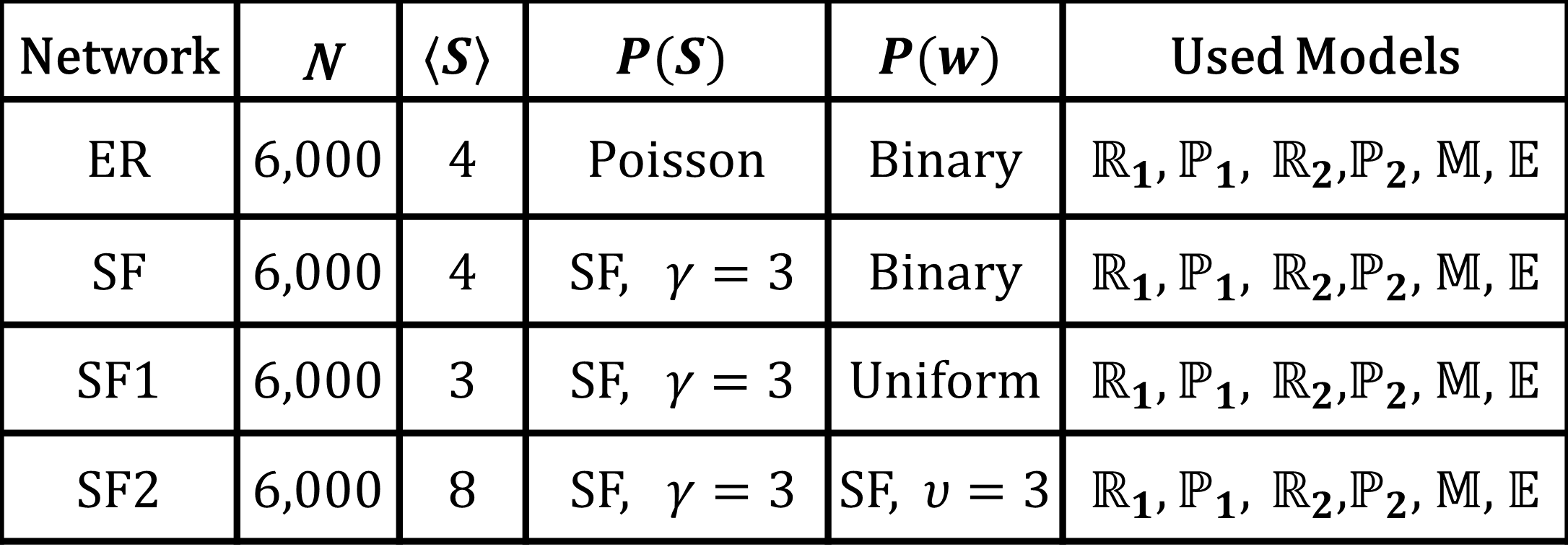}
   	\caption{
   		\textbf{Model networks.} Summary of all model networks used to exemplify our formalism.}
   	\label{Table2}
   \end{table}
   
\noindent {\bf ER}. 
An Erd\H{o}s-R\'{e}nyi random network
with $N = 6,000$ nodes and an average  degree of 
$\langle S \rangle = 4$.

\noindent {\bf SF}. 
A binary scale-free network with $N = 6,000$ nodes,
$\langle S \rangle = 4$ and a degree distribution following
$P(S)\sim S^{-\gamma}$ with  $\gamma=3$, constructed using the 
Barab\'{a}si-Albert model \cite{Albert2002}.
          
\noindent {\bf SF1}. 
Using the underlying topology of SF we added uniformly distributed weights
extracted from $W \sim \mathcal{U}(0.1 - 0.9)$.

\noindent {\bf SF2}. 
Using the underlying topology of SF we extracted the weights $W_{ij}$ 
from a scale-free probability density function
$P(w) \sim w^{-\nu}$ with $\nu = 3$, resulting in an extremely heterogeneous network, featuring a scale-free topology with scale-free weights.

   \begin{table}[h]
 	\centering
 	
  	\includegraphics[width=1\textwidth]{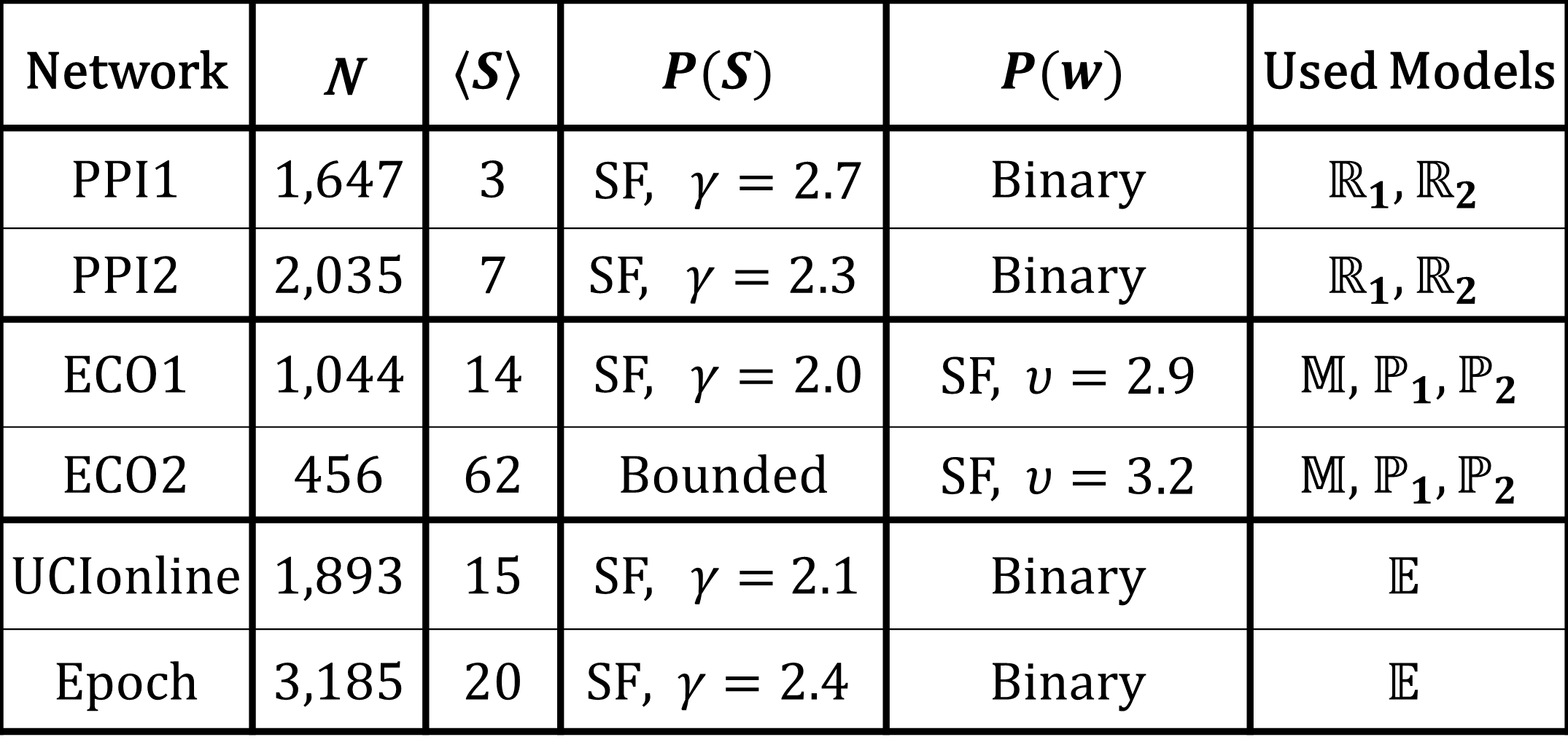}
	\caption{
 	\textbf{Real networks.} We implemented our theory on a set of highly diverse empirical networks, including social, biological and ecological networks. For each network we characterize the weighted degree distribution $P(S)$, bounded versus scale-free (SF), and show the empirically extracted scaling exponent $\gamma$, where relevant. For the weighted networks ECO1/2 we also present the scaling exponent $\nu$ of $P(w)$. On each network we ran the relevant models, {\it e.g.}, epidemic spreading ($\E$) on the social networks.}
  \label{Table3}
      \end{table}

\noindent {\bf UCIonline}. 
An instant messaging network from the University of California Irvine \cite{Opsahl2009}, capturing  
$61,040$ transactions between $1,893$ users during a    
$T = 218$ day period. 
Connecting all individuals who exchanged messages throughout the period, we obtain a network of $1,893$ nodes with $27,670$ links, exhibiting a fat-tailed degree distribution.

\noindent {\bf Email Epoch}. 
This dataset monitors $\sim 3\times 10^5$ emails exchanged between $3,185$
individuals over the course of $T \sim 6$ months \cite{Eckmann2004}, 
giving rise to a scale-free social network with $63,710$ binary links.
         
\noindent {\bf Protein-protein interaction network PPI1}.
The yeast scale-free protein-protein interaction network, consisting of 
$1,647$ nodes (proteins) and $5,036$ undirected links, representing chemical interactions between proteins \cite{Yu2008}. 

\noindent {\bf PPI2}. 
The human protein-protein interaction network, a scale-free network, consisting of $ N = 2,035$ nodes (protein) and $L = 13,806$
protein-protein interaction links \cite{Rual2005}.
    
\noindent {\bf ECO1 and ECO2}.
To construct mutualistic networks we collected data on symbiotic ecological interactions of plants and pollinators in Carlinville Illinois from \cite{EcoNetworks}.
The resulting $456 \times 1,429$ network $\m M ik$ is a bipartite graph linking the $456$ plants with their $1,429$ pollinators. 
When a pair of plants is visited by the same pollinators they mutually benefit each other indirectly, by increasing the pollinator populations. 
Similarly pollinators sharing the same plants also posses an indirect mutualistic interaction.
Hence we can collapse $\m Mik$ to construct two mutualistic networks: 
The $1,429 \times 1,429$ pollinator network ECO1 and the $456 \times 456$ plant network ECO2.
The resulting networks are 

\begin{equation}
B_{kl}=\sum_{i=1}^{456}\dfrac{M_{ik}M_{il}}{\sum_{s=1}^n M_{is}},
\label{def:Mutualistic_weight1}
\end{equation}

\noindent
for the pollinator network (ECO1), and

\begin{equation}
A_{ij}=\sum_{k=1}^{1,429}\dfrac{M_{ik}M_{jk}}{\sum_{s=1}^n M_{sk}},
\label{def:Mutualistic_weight2}
\end{equation}

\noindent
for the plant network (ECO2). 
In both networks the numerator equals to the number of mutual plants ($\m Bkl$) or pollinators ($\m Aij$). 
For each mutual plant $i$ (pollinator $k$) we divide by the overall number of plants (pollinators) that share $i$ ($k$). 
Hence, the weight of the mutualistic interaction in, {\it e.g.}, $\m Aij$ is determined by the density of mutual symbiotic relationships between all plants, where: 
(i) the more mutual pollinators $k$ that plants $i$ and $j$ share the stronger the mutualistic interaction between them; 
(ii) on the other hand the more plants pollinated by $k$ the smaller is its contribution to each plant. A similar logic applies also for the pollinator network $\m Bij$. 
This process potentially allows us to have isolated components, {\it e.g.}, single disconnected nodes. The state of these isolated nodes is decoupled from the state of the rest of the network, and hence in our analysis we only focused on the giant connected component of 
$A_{ij}$ and $B_{ij}$, 
comprising all $456$ plants, rendering $A_{ij}$ to be a fully connected component, but only $1,044$ pollinators, eliminating $385$ isolated pollinators.
\section{Additional results from empirical networks}
\label{Empirical}

\begin{figure}
\centering
\includegraphics[width=1.0\textwidth]{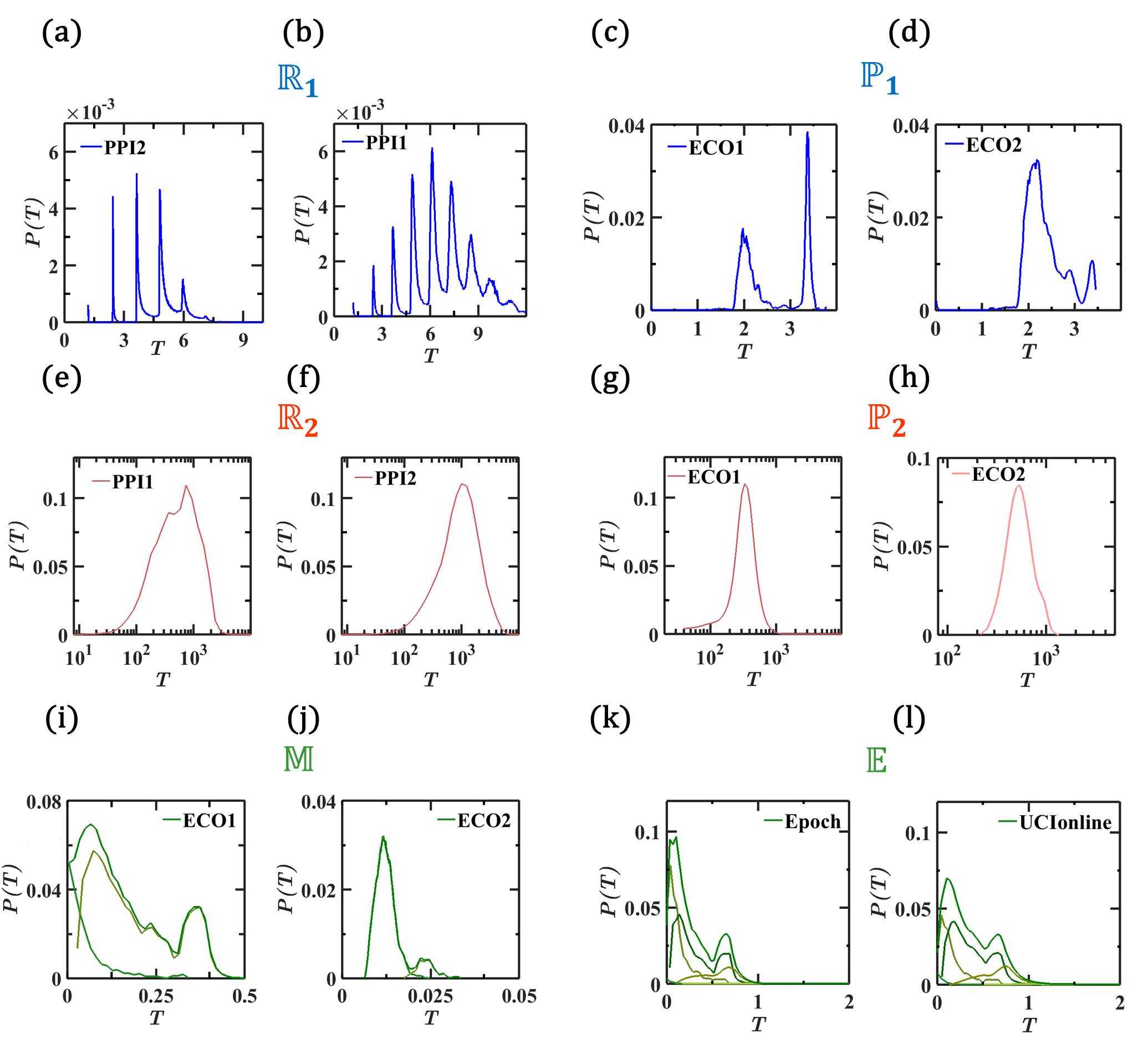}
\caption{
\textbf{Response time density function in empirical networks.} 
(a) - (b)
$P(T)$ vs.\ $T$ as obtained from $\R_1$ on the protein interaction networks PPI1 and PPI2.
(c) - (d)
$P(T)$ vs.\ $T$ as obtained from $\P_1$ on the ecological networks ECO1 and ECO2. Here we observe fewer peaks due to the short paths in these systems. The width of the peaks is driven by the weight distribution, creating variance within each shell of equidistant nodes.
(e) - (h)
$P(T)$ for the degree driven $\R_2$ and $\P_2$ on the same networks as above.
(i) - (l)
In the composite dynamics $\M$ and $\E$, $P(T)$ is characterized by multiple overlapping peaks. The density $P(T|L_{ij})$, capturing the response times within each $L_{ij}$-shell, is also shown (shades of green). ECO2, whose diameter equals $2$ shows only two peaks, as expected. ECO1, Epoch and UCIonline exhibit each two or three peaks within each shell - a consequence of the composite dynamics, in which low degree nodes respond at later times.  
}
\label{FigPT1}
\end{figure}

To complement the results presented in the main paper we include here observations extracted from our set of empirical networks (Sec.\ \ref{Networks}), comprising $12$ combinations of networks and dynamics, as appear in Table \ref{Table3}. The scaling relationship $\tau_i \sim S_i^{\theta}$ has already been tested in the main text (Fig.\ 4) on all $36$ systems, including our empirical networks, hence we focus below on the structure of $P(T)$ and the layouts predicted by our universal metric $\L ij$ (Eq.\ (5) in main text).

\subsection{Distance driven propagation} 

In Fig.\ \ref{FigPT1}a - d we show $P(T)$ vs.\ $T$ (blue) as obtained for the distance driven $\R_1$ and $\P_1$ implemented on PPI1 and PPI2 ($\R_1$) and on ECO1 and ECO2 ($\P_1$). As predicted for these distance driven dynamics, $P(T)$ exhibits separated sharp peaks, corresponding to the discrete lengths of all paths $L_{ij}$. The effect is clearly pronounces in PPI1 and PPI2, and slightly less sharp in ECO1 and ECO2. The reason is that these ecological networks have distributed weights (Table \ref{Table3}), and hence at each distance, we observe some level of variance in the response times, a consequence of the weight heterogeneity along all paths, which is reflected in $P(T)$ by the width of the observed peaks. Also note, that ECO1 and ECO2 are rater dense, and therefore have only two or three shells, with $\max(L_{ij})$ being only $2$ for ECO2 and $3$ for ECO1, hence the small number of peaks for these systems. 

\begin{figure}
\centering
\includegraphics[width=0.8\textwidth]{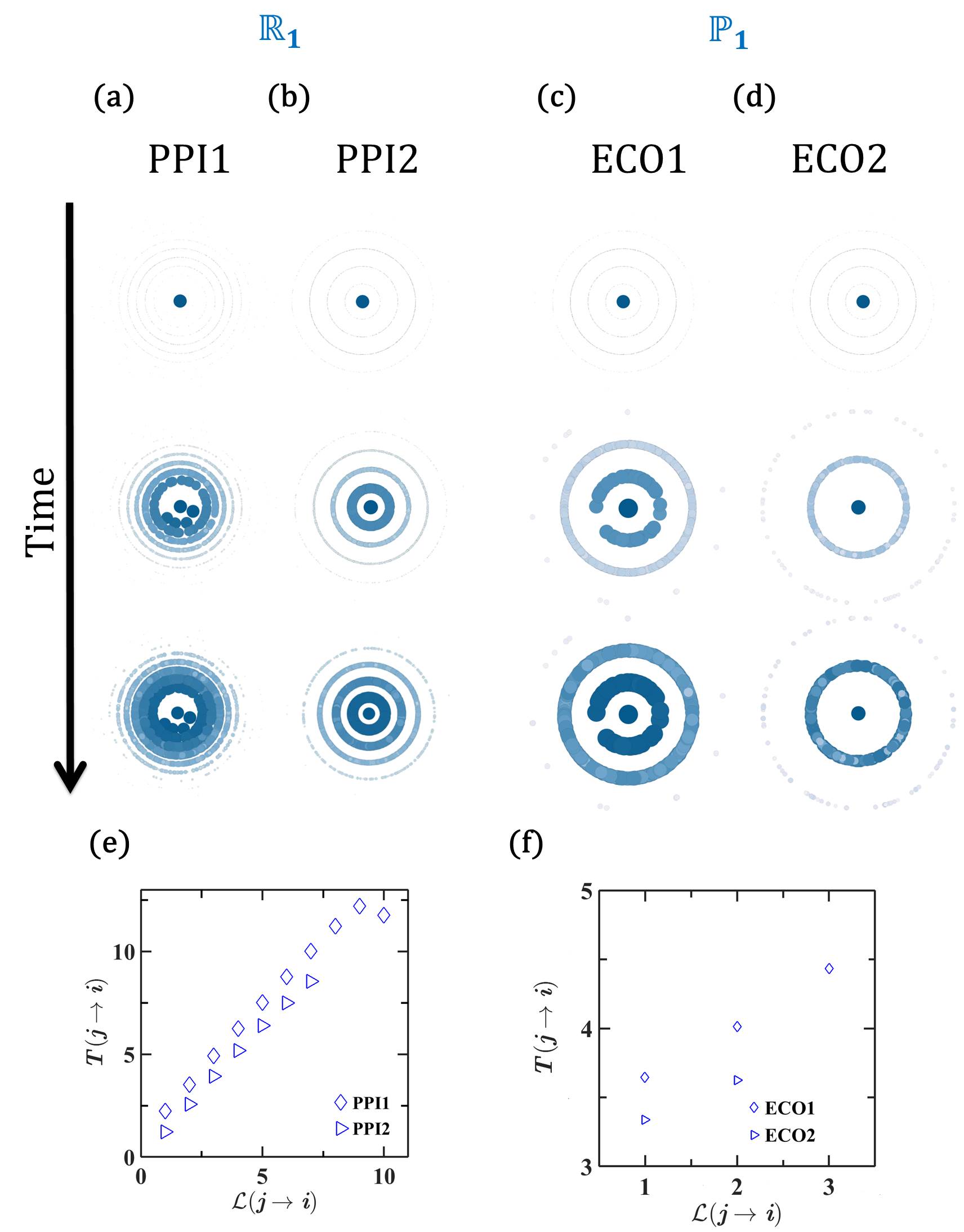}
\caption{
\textbf{The universal distance $\L ij$ in empirical networks.} 
Results obtained from our four distance driven systems. 
}
\label{FigL1}
\end{figure}

We further tested our universal dynamic metric $\L ij$, as predicted in Eq.\ (5) of the main text. The results, presented in Fig.\ \ref{FigL1} confirm that indeed, these four systems all exhibit distance driven propagation, expressed through the discrete shells characterizing the traveling signals. For ECO1 and ECO2 we only observe $2$ or $3$ shells, due, again, to the relatively small diameter of these systems.     

\subsection{Degree driven propagation}

\begin{figure}
\centering
\includegraphics[width=0.8\textwidth]{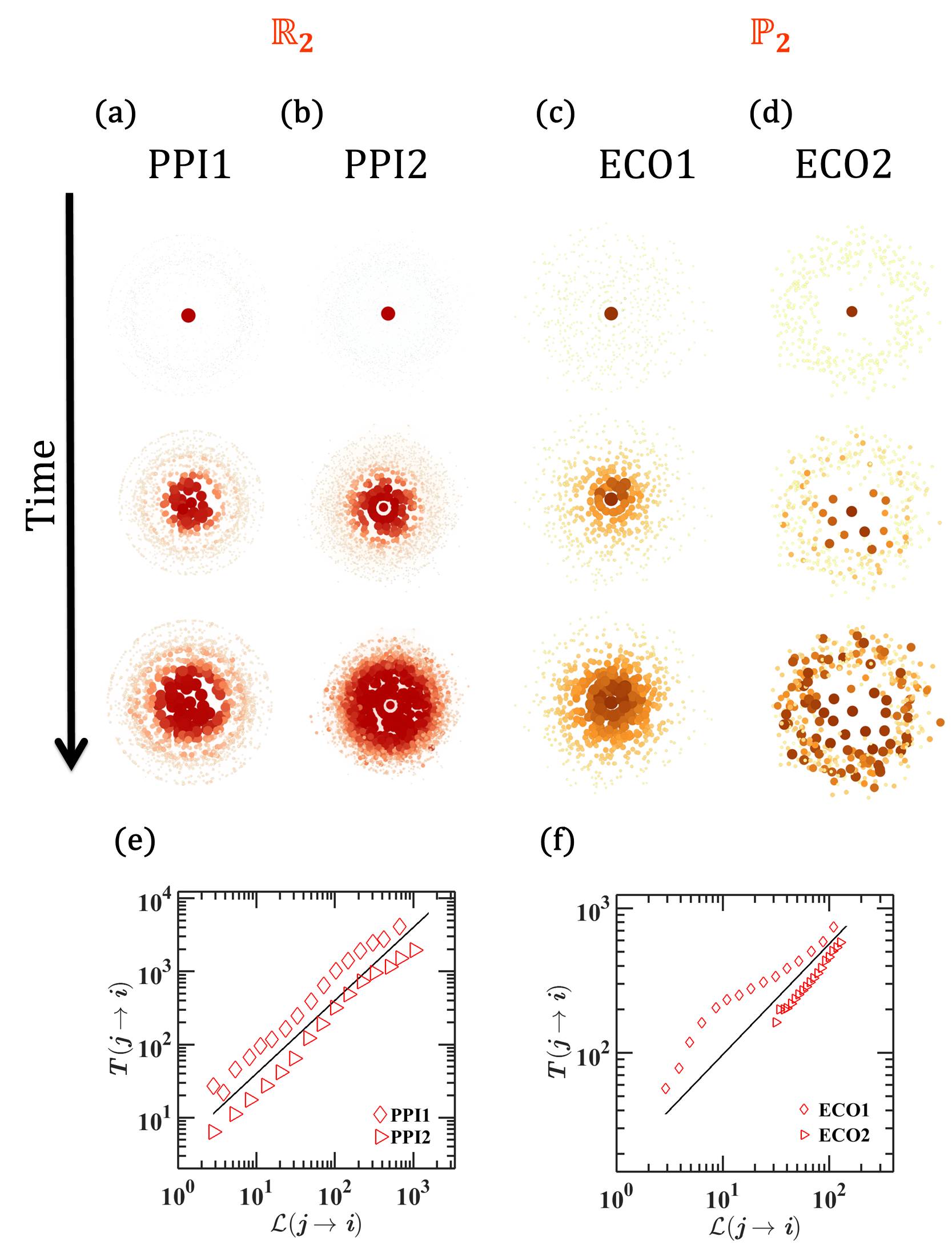}
\caption{
\textbf{The universal distance $\L ij$ in empirical networks.} 
Results obtained from our four degree driven systems. 
}
\label{FigL2}
\end{figure}

Our testing ground includes four degree driven systems: $\R_2$ combined with PPI1/2 and $\P_2$ combined with ECO1/2. The density $P(T)$ for these systems is presented in Fig.\ \ref{FigPT1}e - h (red), following precisely the anticipated form, as predicted and observed on the model networks of Fig.\ 3 of the main text. The spatio-temporal layout, $\L ij$, for these four systems appears in Fig.\ \ref{FigL2}. While the results for PPI1/2 and ECO1 follow our predictions with high accuracy, we find that for ECO2, the $\L ij$ prediction exhibits rather high levels of noise (Fig.\ \ref{FigL2}d). Still, the average propagation is well approximated by $\L ij$ (Fig.\ \ref{FigL2}f). Indeed, ECO2, a small ($N = 456$) and relatively dense ($\langle S \rangle = 62$) network, is characteirzed by many loops ($C \approx 0.1$), and extremely short paths ($\max(L_{ij}) = 2$), and hence does not adhere to our model assumptions. This has little effect on our macroscopic predictions, $\theta$, $P(T)$, average $\L ij$ vs.\ $\T ij$, but does impact the quality of the more node-specific layouts of Fig.\ \ref{FigL2}d.  

\subsection{Composite propagation}

\begin{figure}
\centering
\includegraphics[width=1.0\textwidth]{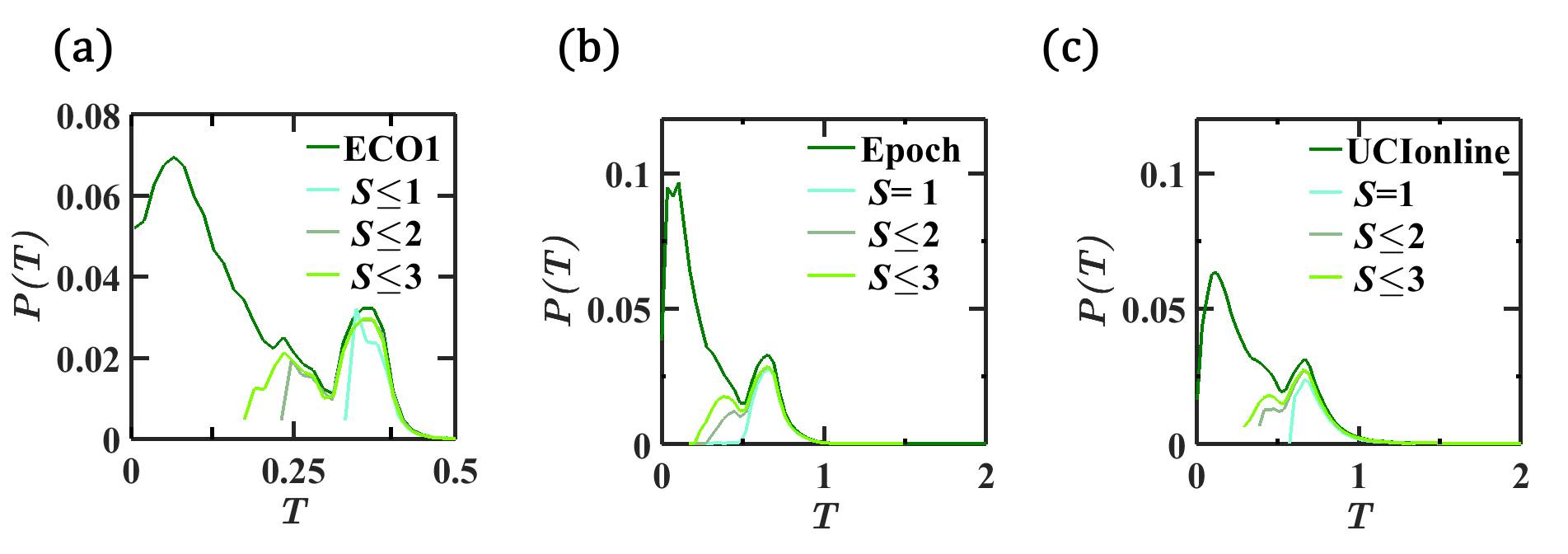}
\caption{
\textbf{A close up on the composite structure of $P(T)$.} 
We focus on $P(T)$ vs.\ $T$ on (a) $\M$ on ECO1; (b) $\E$ on Epoch and (c) $\E$ on UCIonline. The different peaks, in these systems, correspond to the low degree target nodes, which respond late (large $T$) under composite dynamics. This structure of $P(T)$ exposes the interplay of distance and degrees, characterizing the composite universality class. 
}
\label{FigPT3}
\end{figure}

\begin{figure}
\centering
\includegraphics[width=0.8\textwidth]{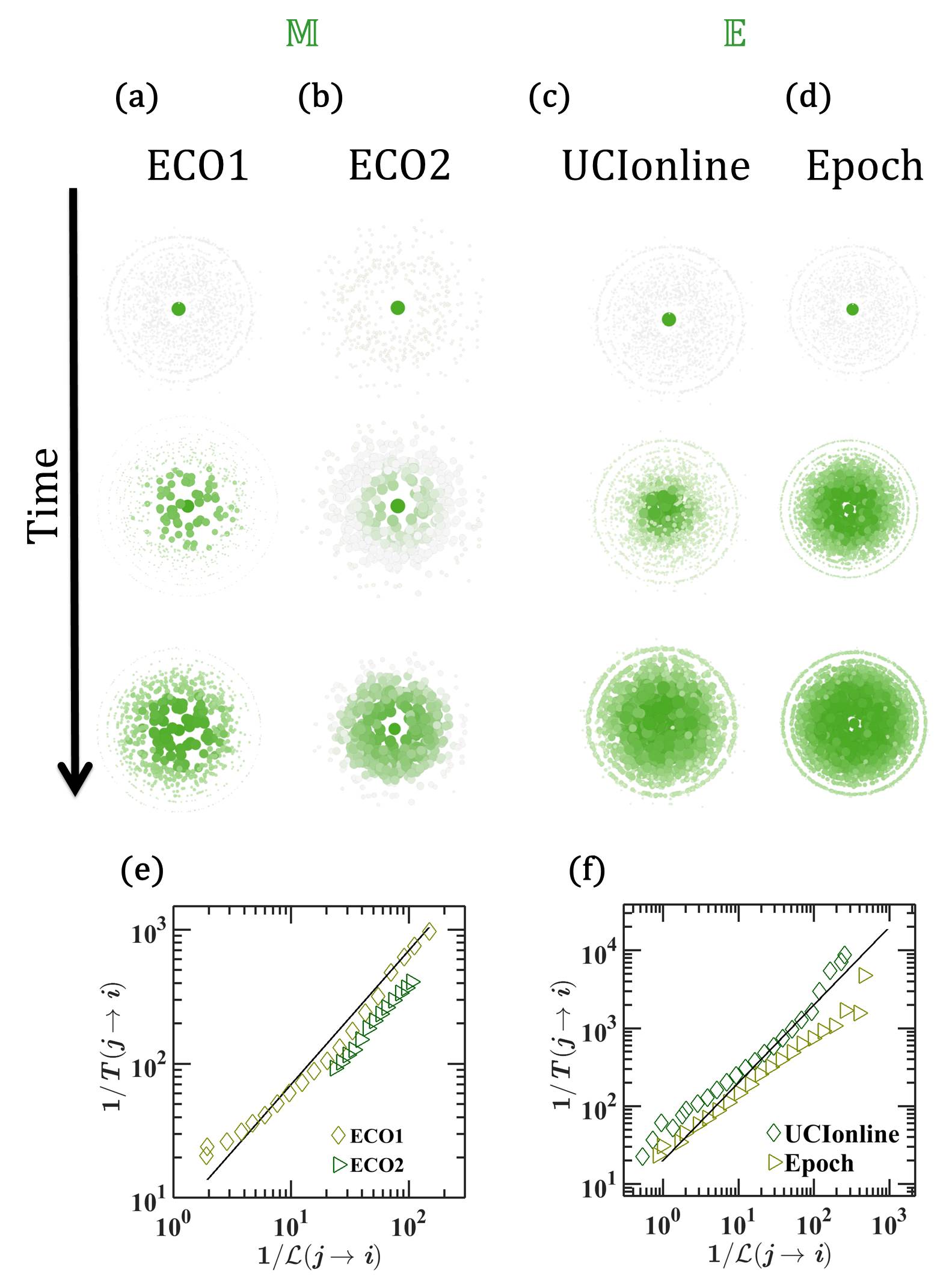}
\caption{
\textbf{The universal distance $\L ij$ in empirical networks.} 
Results obtained from our four composite systems. 
}
\label{FigL3}
\end{figure}

Our four empirical systems in the composite class include $\M$, applied to ECO1 and ECO2, and $\E$, applied to UCIonline and Epoch. In Fig.\ \ref{FigPT1}i - l we show $P(T)$ (green), as obtained from these four systems. As predicted, we find multiple overlapping peaks - the fingerprint of the composite dynamic class. Interestingly, in these empirical settings the composite interplay between $L_{ij}$ and $P(S)$ is more complex that that observed on the model networks. For instance in ECO1, Epoch and UCIonline, the inner functions representing $P(T|L_{ij})$ (shades of green) indeed show the anticipated effect of network distance, with the progression of the inner peaks as $L_{ij}$ is increased. However, these three systems also feature secondary peaks within the same shell, \textit{i.e.} $P(T)$ within $L_{ij} = 2$ (dark green) is bi-modal, showing that within the same distance, we observe two typical response times. This is a direct consequence of the composite dynamics, in which $P(T)$ depends both on $L_{ij}$ and on $S_i$. To observe this we focus on these three systems in Fig.\ \ref{FigPT3}, this time showing $P(T|S_i)$, the $\T ij$ density of target nodes with a given degree $S_i$. We find, indeed, that the secondary peaks are driven by the low degree nodes within each shell, whose response time is large, due to their low weighted degree ($\theta < 0$). This illustrates the essence of the composite class, where $\T ij$ is determined both by the $L_{ij}$-shells, but also by the distribution of $S_i$ within each shell, leading, in the case of these empirical networks, to such non-trivial structure of $P(T)$. The spatio-temporal layouts for these systems appear in Fig.\ \ref{FigL3}.

\clearpage 
\section{Additional validation}
\label{Pertubation and Degree Correlation}

Our analytical derivations, outlined in Secs.\ \ref{Transient} and \ref{Models} are exact under 
two main assumptions: (i) the perturbative limit of small signals $\dif x$, which allows us to use linear response theoretic tools; (ii) the configuration model \cite{Newman2010} pertaining to $A_{ij}$, according to which node $i$'s nearest neighbor statistics are independent of $i$. In real scenarios we are often confronted by large perturbations, or by empirical networks, which may violate, to some extent, the \textit{clean} picture of the configuration model. Therefore we tested the robustness of our analytically predicted scaling, (\ref{tauk}), against deviations from assumptions (i) and (ii) above. Specifically, regarding (i), we test the impact of large perturbations, ranging from $10\%$ to $100\%$, culminating in full node knockout. Regarding (ii) we introduce two topological features that are frequently observed in real networks, but violate the configuration model framework: degree-degree correlations \cite{Newman2002} and clustering. These non-local topological characteristics are a fingerprint of non-random connectivity, overriding the essential ingredient of the configuration model. 

\subsection{The effect of large perturbations}

\begin{figure}[h!]
\centering
\includegraphics[width=0.6\textwidth]{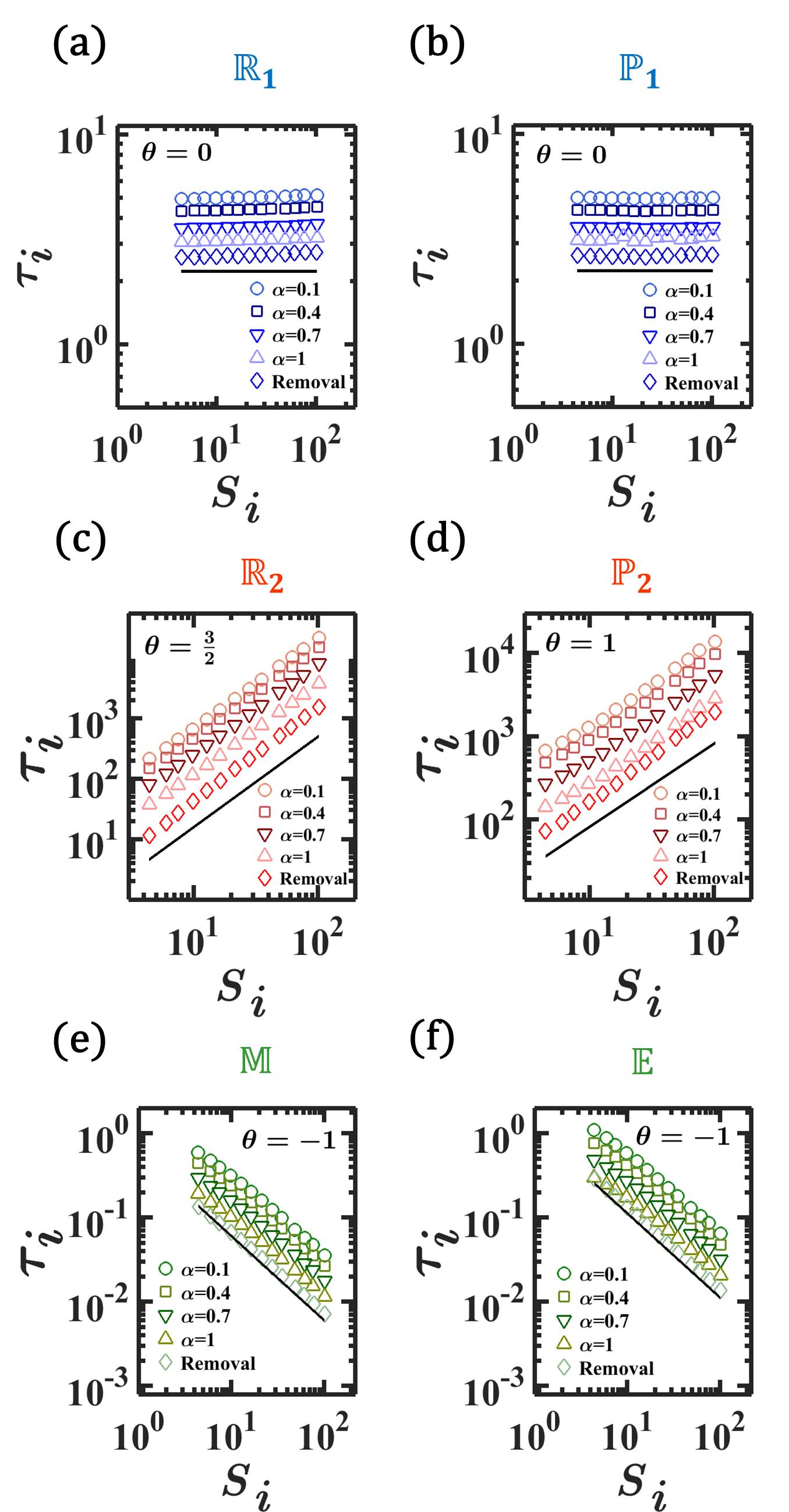}
\caption{
\textbf{The effect of perturbation size}. To test the limits of our linear response framework we measured $\tau_i$ vs.\ $S_i$, as obtained for large signals, representing an $\alpha = 10\%$ (circles), $40\%$ (squares), $70\%$ (down-triangles) and $100\%$ (up-triangles) perturbation. We also tested the scaling under complete node knockout (Removal, diamonds). We find that perturbation size has no visible effect on the macroscopic patterns of flow, with $\theta$ consistently adhering to the theoretically predicted value (solid lines).  	 
	}
	\label{Figalpha}
\end{figure}

Measuring $\T ij$ entails introducing a signal, $\Delta x_j$, to the steady state activity $x_j$ of the source node $j$, and observing the flow of information as it propagates from $j$ to $i$. In our derivations we resort to the perturbative limit ($\Delta x_j \rightarrow \dif x_j$), where $\alpha = \dif x_j / x_j \ll 1$, a small perturbation, that allows us to use linearization to achieve analytical advances. Specifically, in our numerical experiments we set the magnitude of our signals to $10\%$ of the source's steady state, namely $\alpha = 0.1$. In Fig.\ \ref{Figalpha} we examine the impact of larger perturbations, setting $\alpha = 0.4$ (squares), a $40\%$ perturbation, $\alpha = 0.7$ (down-triangles), a larger perturbation of $70\%$, and even $\alpha = 1$ (up-triangles) a signal of the same size as the node's unperturbed activity. We find that the predicted scaling $\theta$ is extremely robust, with the size of the perturbation having no visible effect. We further tested information propagation under the full knockout of the source node, namely removing node $j$ and observing the spatio-temporal system response (diamonds). Such node removal represents a common procedure to observe sub-cellular dynamics via controlled genetic knockouts \cite{Kauffman2004}. It also arises in naturally occurring settings, such as in spontaneous component failure in \textit{e.g.}, the power grid \cite{Zhao2016}. We find that even under these extreme conditions our predicted scaling remains valid, indicating that our predictions are highly robust against perturbation size.   

This lack of sensitivity is rooted in the well-established robustness of scaling relationships, which are often unaffected by small deviations and discrepancies \cite{Wilson1975}. This is especially relevant in a network environment, where local perturbations rapidly decay (exponentially) as they penetrate the network \cite{Barzel2013}. Under these conditions even a large local perturbations will have only a small effect on all individual nodes in its vicinity. Therefore, the consequent responses of the signal's direct neighbors, next neighbors and so on, can be well-approximated by the perturbative limit, even if the original $j$-signal was in violation of this limit. Hence we find that the linear response framework remains valid even under unambiguously \textit{large} perturbations. 

\subsection{The effect of clustering}

Next we consider the impact of clustering $C$, representing the network's tendency to from triads, in which there is an increased probability for an $n,m$ link, if $n$ and $m$ share a mutual neighbor $i$. Under the configuration model assumption, clustering tends to zero if the network is sparse and $N \rightarrow \infty$ \cite{Newman2010}. Most empirical networks, however, feature non vanishing levels of clustering, in some cases reaching an order of $C \sim 10^{-1}$ \cite{Barabasi2002}, significantly higher than that expected in a random connectivity. To measure node $i$'s clustering we write  

\begin{equation}
C_i = \dfrac{\displaystyle \sum_{m,n = 1}^N A_{im} A_{in} A_{nm}}{\dbinom{k_i}{2}},
\label{def:clustering_coefficient}
\end{equation}

\noindent
in which the numerator counts the number of \textit{actual} triads involving nearest neighbors of $i$, and the denominator equals to the number of \textit{possible} triads around $i$, \textit{i.e.} the number of potential pairs among $i$'s $k_i$ nearest neighbors. Hence $0 \le C_i \le 1$ is the fraction of potential triads that are actually present among $i$'s neighbors. The clustering of the \textit{network} is then obtained by averaging over all nodes as

\begin{equation}
C = \dfrac{1}{N} \sum_{i = 1}^{N} C_i.
\label{Clustering}
\end{equation}

\begin{figure}[h!]
\centering
\includegraphics[width=0.6\textwidth]{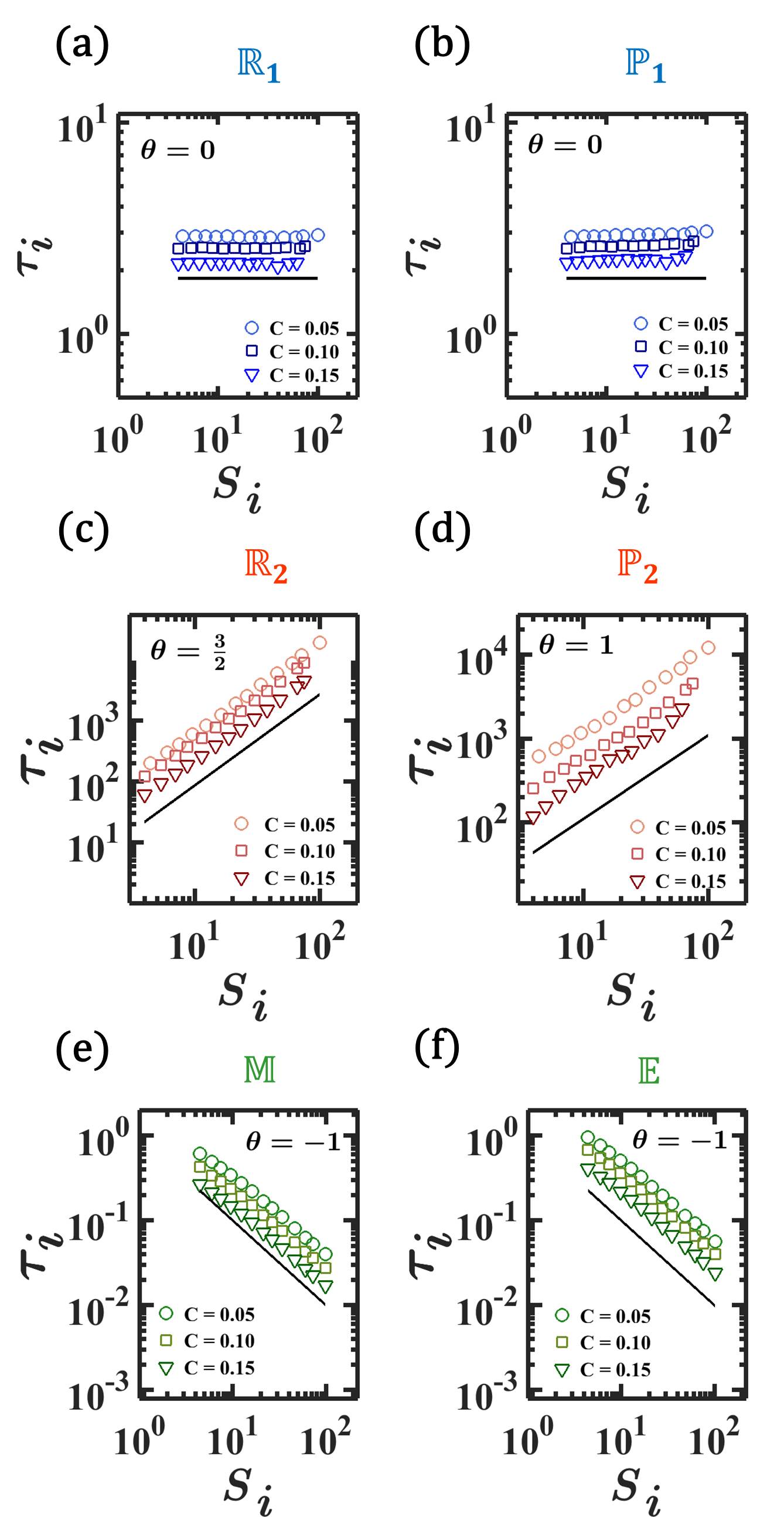}
\caption{
\textbf{\bf The impact of clustering $C$}.
$\tau_i$ vs.\ $S_i$ as obtained from SFC1 - SFC3, featuring increasing levels of clustering $C = 0.05$ to $0.15$. Despite the clustering the predicted scaling in each dynamics ($\theta$, solid lines) remains valid. 	 
}
\label{FigClustering}
\end{figure}

\noindent
In Table \ref{Table4} we show the clustering $C$ as obtained from our set of empirical networks. We find that for some of these networks $C$ is rather high, in some cases reaching as much as $C = 0.2567$ (Epoch). Still, as demonstrated in the main text, our analytical predictions performed well, even under these challenging conditions of extreme clustering. This indicates that our predictions are robust against empirically observed levels of clustering. To further examine the effects of clustering in a controlled fashion, we used the scale-free network SF, and gradually rewired it to increase its clustering to $C_1 = 0.05, C_2 = 0.1$ and $C_3 = 0.15$ (Table \ref{Table4}), generating three model networks, SFC1 - SFC3, with controlled levels of clustering. We then measured $\tau_i$ vs.\ $S_i$ on each of these networks. We find again that even extreme levels of clustering ($C = 0.15$ is two orders of magnitude higher than the configuration model expectation value), our theoretical predictions are consistently sustained (Fig.\ \ref{FigClustering}). 

\subsection{The effect of degree-degree correlations}

\begin{figure}[h!]
\centering
\includegraphics[width=0.6\textwidth]{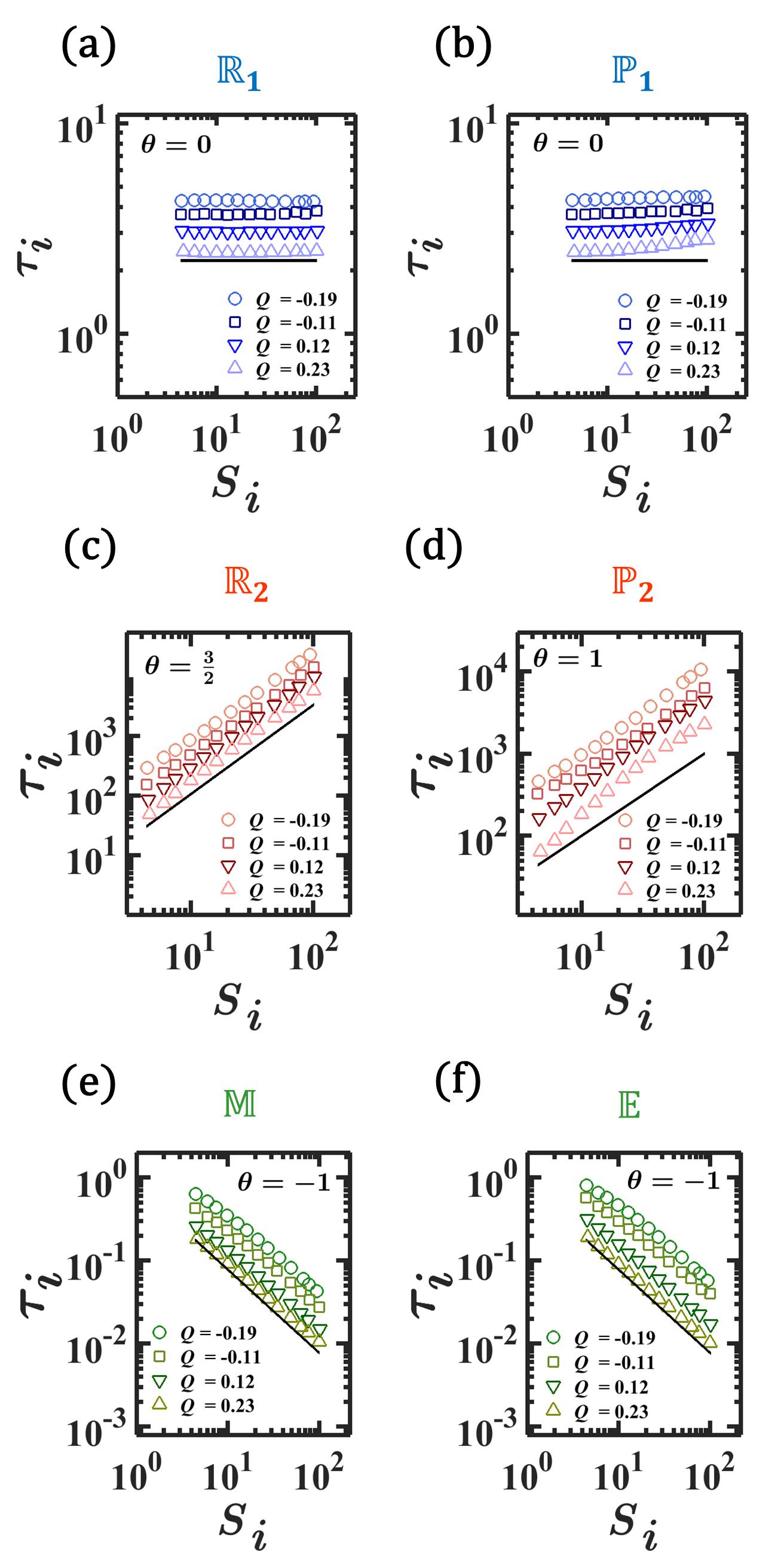}
\caption{
\textbf{
\bf The impact of degree correlations $Q$}.
$\tau_i$ vs.\ $S_i$ as obtained from SFQ1 - SFQ4, featuring negative and positive degree correlations. Despite these correlation levels the predicted scaling in each dynamics ($\theta$, solid lines) remains valid. 	 	 
	}
	\label{FigDegreecorrelation}
\end{figure}

As our final test, we examine the effect of degree correlations $Q$, as defined in Ref.\ \cite{Newman2002}. As before, we first observe the correlation levels exhibited by our set of empirical networks, finding that they feature rather high levels of degree correlations (Table \ref{Table4}). The fact that our predictions cover these networks is, as before, an indication of our theory's robustness against empirically observed correlations. To complement this finding we rewired SF, once again, this time to exhibit increasing levels of positive and negative degree correlations, producing SFQ1 - SFQ4, as detailed in Table \ref{Table4}. As in the case of clustering, the results, presented in Fig.\ \ref{FigDegreecorrelation}, show that our predictions are largely unharmed by $Q$, indicating their low sensitivity to the configuration model assumption of Sec.\ \ref{Transient}.

\begin{Frame}
\textbf{Robustness of predicted universality classes}. 
Our theory provides both quantitative as well as qualitative predictions. At the quantitative level, we predict the precise value of $\theta$, allowing us to provide the precise response times of all nodes (Fig.\ 4 in main text). No less important are, however, our qualitative predictions, that allow us to translate $\theta$ into direct insights on the macroscopic propagation patterns of a networked system. This is observed by the distinct structures of $P(T)$ (Fig.\ 3g - l in main text), the different roles of network paths $L_{ij}$ (Fig.\ 3m - r in main text), and the class-specific contribution of $P(S)$ (Fig.\ 5 in main text). All of these observations represent macro-level dynamic patterns that determine how the \textit{system} (as opposed to specific nodes) manifests information propagation. Such intrinsic characteristics are seldom sensitive to microscopic discrepancies.

We further argue that even if the precise value of $\theta$ deviates due to some specific departures from our model assumptions - deviations that we have not observed in our extensive numerical tests - still, the implications on the macro-scale behavior of the system, indeed, the \textit{qualitative} insight that our theory aims to provide, will ultimately be marginal. For instance, consider a deviation in one of our dynamics, say the degree driven $\R_2$, which under some hypothetical conditions features, \textit{e.g.}, a decrease in its observed $\theta$ from the theoretically predicted $\theta = 3/2$ to, say, $\theta \approx 1$. This may constitute a significant discrepancy in terms of our quantitatively predicted scaling, but will not significantly impact the observed propagation patterns, which will remain within the degree-driven class. Indeed, micro, or even meso-scopic discrepancies from our model assumptions cannot cause a qualitative shift to a different class, turning, for instance from degree-driven to distance-driven or composite. Such transition can only be done by altering the system's internal mechanisms, such as shifting from $\R_2$ ($\theta = 3/2$) to $\E$ ($\theta = -1$), a change in the \textit{physics} of the node interactions, which requires a fundamental intervention, unattainable by minor discrepancies.     
\end{Frame}

 \begin{table}[h]
\centering
\includegraphics[width=1\textwidth]{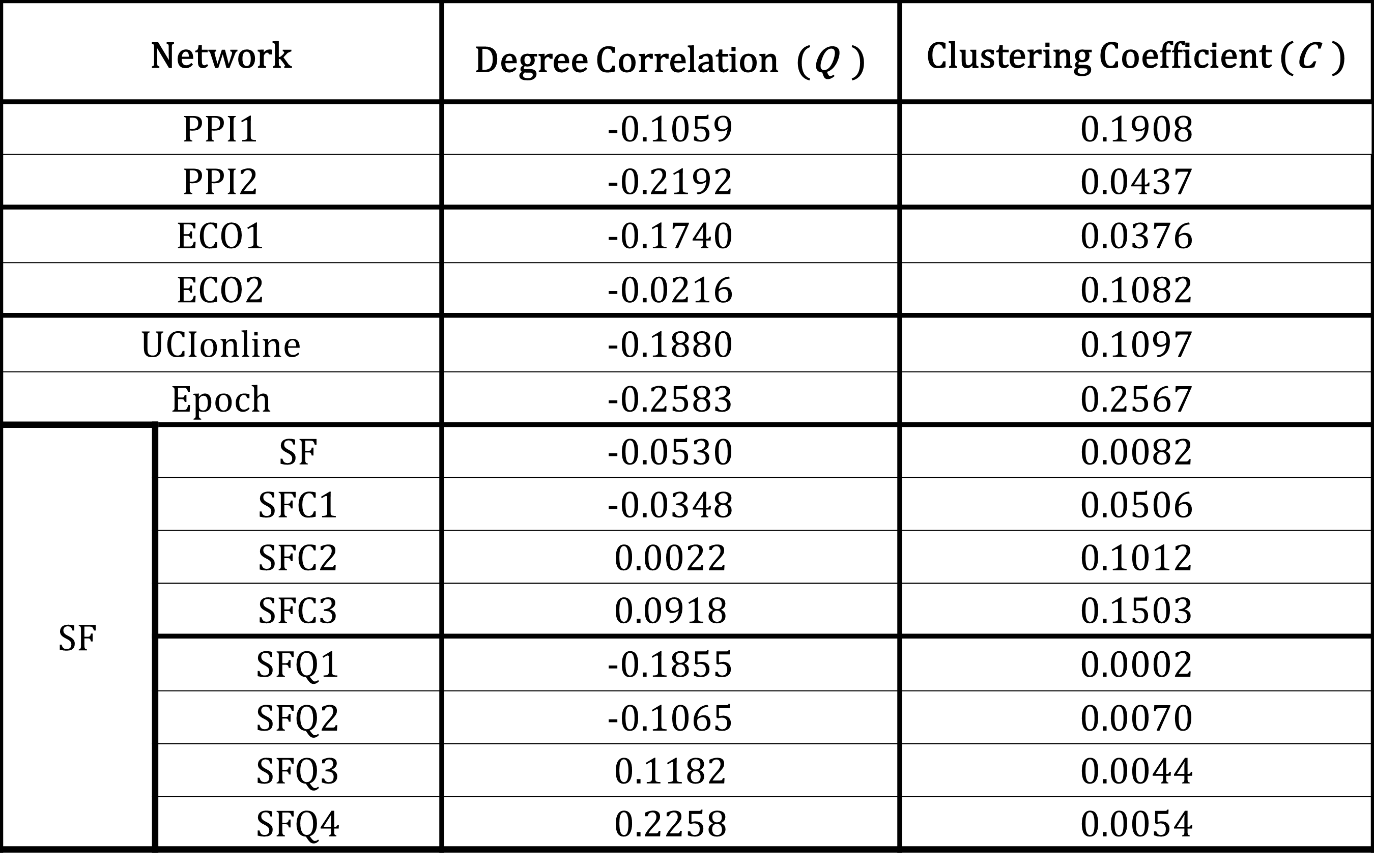}
\caption{
\textbf{Degree correlations ($Q$) and clustering ($C$) of our model and empirical networks.} 
We measured $Q$ and $C$ from our set of empirical networks. Our results seem to have been unaffected by the high levels of $Q, C$. We also rewired our model scale-free network SF to increase its clustering (SFC1 - SFC3) and degree-correlations (SFQ1 - SFQ4) in a controlled fashion.}
\label{Table4}
\end{table}

\clearpage

\bibliographystyle{unsrt}
\bibliography{Bibliography}


\clearpage